\documentclass[smallextended]{svjour3}
\usepackage[table,dvipsnames]{xcolor}
\usepackage{colortbl}
\usepackage{tabularx}
\usepackage{array}
\usepackage{dashrule}
\usepackage{multirow}
\usepackage{enumitem}
\usepackage[framemethod=TikZ]{mdframed}
\usepackage{changepage}
\usepackage{listings}
\usepackage{graphicx}
\usepackage{float}
\usepackage{natbib}
\usepackage{printlen}
\usepackage{longtable}
\usepackage{pdflscape}
\usepackage{booktabs}
\usepackage{arydshln}
\usepackage{amssymb}
\usepackage{adjustbox}
\usepackage[hidelinks]{hyperref}
\usepackage{makecell}
\usepackage{lmodern}
\usepackage{xspace}
\usepackage{rotating}
\usepackage{booktabs}
\usepackage{array}
\usepackage[normalem]{ulem}
\usepackage{siunitx}
\usepackage{pifont}
\newcolumntype{C}[1]{>{\centering\arraybackslash}p{#1}}

\definecolor{codegreen}{rgb}{0,0.6,0}
\definecolor{codegray}{rgb}{0.5,0.5,0.5}
\definecolor{codepurple}{rgb}{0.58,0,0.82}
\definecolor{backcolour}{rgb}{0.95,0.95,0.92}
\definecolor{mutedblue}{RGB}{80,100,140}
\lstdefinestyle{mystyle}{
    commentstyle=\color{codegray},
    keywordstyle=\color{blue},
    numberstyle=\scriptsize,
    stringstyle=\color{codegreen},
    basicstyle=\ttfamily\scriptsize,
    breakatwhitespace=false,
    breaklines=true,
    captionpos=t,
    keepspaces=true,
    numbersep=5pt,
    showspaces=false,
    showstringspaces=false,
    showtabs=false,
    tabsize=2,
    xleftmargin=0.4ex,
    linewidth=\linewidth,
    language=python,
    escapeinside={(*@}{@*)}
}

\usepackage{ulem}
\usepackage{stfloats}
\fnbelowfloat

\newcommand{\circled}[1]{\raisebox{.5pt}{\textcircled{\raisebox{-.9pt} {#1}}}}

\newcommand{\ie}{\textit{i.e.,}\xspace}
\newcommand{\eg}{\textit{e.g.,}\xspace}

\newcommand{\secref}[1]{Section~\ref{#1}\xspace}

\newcommand{\figref}[1]{Fig.~\ref{#1}\xspace}
\newcommand{\listref}[1]{Listing~\ref{#1}\xspace}
\newcommand{\tabref}[1]{Table~\ref{#1}\xspace}
\newcommand{\etal}{\textit{et al.}\xspace}
\newcommand{\solo}{\textit{solo-coding}\xspace}
\newcommand{\collaborative}{\textit{collaborative}\xspace}
\newcommand{\fullyautomated}{\textit{fully-automated}\xspace}
\newcommand{\agentic}{\textit{agentic}\xspace}

\newcommand{\TPR}{TPR\xspace}
\newcommand{\SCB}{SCB\xspace}
\newcommand{\BCB}{BCB\xspace}
\newcommand{\SC}{SC\xspace}
\newcommand{\BC}{BC\xspace}
\newcommand{\MCC}{MCC\xspace}
\newcommand{\LCS}{LCS\xspace}
\newcommand{\LED}{LED\xspace}
\newcommand{\TAM}{TAM\xspace}
\newcommand{\TM}{TM\xspace}
\newcommand{\TLOC}{TLOC\xspace}
\newcommand{\CLOC}{CLOC\xspace}
\newcommand{\mIterations}{\# of Iterations\xspace}
\newcommand{\Time}{Time\xspace}

\begin{document}

\title{Vibe Coding: An Experiment with Test-Driven Development}

\author{Moritz Mock \and Barbara Russo}

\institute{Moritz Mock \at Faculty of Engineering, \\Free University of Bozen-Bolzano, \\Bolzano, Italy\\
\email{momock@unibz.it} \and Barbara Russo \at Free University of Bozen-Bolzano, \\Bolzano, Italy\\\email{brusso@unibz.it} }

\date{Received: date / Accepted: date}

\maketitle

\begin{abstract}
\textit{Context:} Conversational Large Language Models (CLLMs) can automatically generate code by collaborating with users through natural language. However, poor collaboration can lead to poor quality output. 

\noindent\textit{Objective:} This exploratory study aims to investigate how humans and CLLMs can collaborate as peers through vibe coding, an approach that integrates principles from prompt engineering, agile design, and human-AI co-creation to enhance collaboration.

\noindent\textit{Method:} We designed four interaction models representing different collaboration patterns in the software development process: the solo model (human-only development), the collaborative model (human–CLLM collaboration), the fully automated model (development autonomously performed by a CLLM), and the agentic model (development autonomously performed by the MetaGPT~X platform). Based on these models, we implemented corresponding Test-Driven Development (TDD) workflows using structured prompts and Python scripts. We then conducted a controlled pre-experimental study with TDD professionals to compare the solo and collaborative workflows. In addition, we performed repeated exploratory executions of fully automated and agentic workflows on the same development tasks to obtain complementary evidence.

\noindent\textit{Results:} Our findings suggest that the choice of interaction model should depend on the development objective. Agentic workflows are best suited for rapid development and functionally correct production code but may introduce additional implementation complexity. However, they may also introduce additional implementation decisions that are not explicitly required by the functional specifications, resulting in untested decision points. In contrast, collaborative workflows produce higher-quality, better-organized test suites. 

\noindent\textit{Conclusions:} Our work explored how humans can collaborate with CLLMs in generating code in highly intensive and demanding development process. Our results also show that accurately reproducing development process workflows is yet an open problem, also in the new agentic technologies.   

\keywords{Conversational Large Language Model \and Vibe Coding \and Human-AI collaboration \and Prompt Engineering \and Test Driven Development}
\end{abstract}

\maketitle

\section{Introduction}
\label{sec:introduction}
Conversational LLMs (CLLMs), such as ChatGPT, are Large Language Models (LLMs) designed to engage in natural, human-like conversations \citep{White2023}. CLLMs have been introduced in software development to streamline and automate various tasks, such as requirement analysis and code generation~\citep{White2024}. As such, the interest of academia and industry in these models has increased rapidly~\citep{Ebert2023}. 
However, introducing CLLMs in software development comes with a cost. 
The quality of the answers obtained may not have the desired level~\citep{ERC2023,ButlerEtAl2025}. The characteristics of the selected model, the data used in model pre-training, and the way developers query it may indeed impact the quality of the generated output~\citep{GoodfellowEtAl2016,BishopBishop2023,Liu2024tse,TieEtAl2024}. Inadequate queries can, for example, lead to unfaithful responses - called hallucinations,~\citep{VaidyaAsif2023,MaynezEtAl2020} - that developers can directly incorporate into their software without further verification~\citep{Fan2023,YangEtAl2024}. The role of human developers and their interaction with an AI is also a key concern~\citep{ButlerEtAl2025,UlfsnesEtAl2024}. Although machines are typically used to replace human developers in automating simple repetitive engineering tasks~\citep{ButlerEtAl2025,UlfsnesEtAl2024,engIT}, developers may still fear that their role could lose value.

Recent literature has explored ways to integrate CLLMs into software development. Among the approaches, vibe and agentic coding are the most popular. The former emphasises co-creation, where developers guide CLLM step by step through prompts, while the latter enables CLLM agents to autonomously plan, execute, test, and iterate on development tasks with minimal human intervention, which is typically limited to setting overall goals~\citep{TholanderJonsson2026}.
These represent two different coding philosophies: vibe coding fosters AI-human collaboration and empowers human creativity through prompting, while agentic coding aims at CLLM autonomy and delegation, allowing humans to define coding objectives~\citep{SapkotaEtAl2025VibeAgentic}.

The goal of our work is to \textit{examine how CLLMs
can be employed in software development}. To this end, we explored human-CLLM collaboration in multiple settings and levels of autonomy. Our main contributions are then summarised as follows:

\begin{itemize}
\item we investigated the participation of CLLMs in software development under varying degrees of autonomy, ranging from vibe coding to agentic coding;
\item we designed and integrated human-CLLM interaction models;
\item we developed novel prompt patterns to support test and code generation;
\item we designed and conducted a controlled pre-experimental study with distributed professional developers on the with Test-Driven Development (TDD) process,~\citep{CampbellEtAl2015};
\item we designed and performed an experiment with CLLM-based fully automated and agentic solutions;
\item we developed a reproducible toolbox to replicate our experiment with professionals in distributed collaborative environments (\eg Google Colab).
\end{itemize}

The remainder of the paper is organised as follows:
\secref{sec:background} provides background on CLLMs, prompt engineering, vibe and agentic coding, and introduces to TDD.
\secref{sec:methodology} presents our methodology for automated TDD with a CLLM, detailing the interaction models we have defined, the prompt design, and the automated vibe coding TDD models.
In \secref{sec:experiment}, we present the research questions and the empirical study.
Followed by \secref{sec:experimentResults}, which presents the results of the empirical study, going over to the discussion and the implications of this work in \secref{sec:discussion}.
\secref{sec:threatValidity} reports threats to the validity of our results.
\secref{sec:relatedWork} discusses previous research related to our work.
Finally, \secref{sec:conclusion} concludes the paper and provides an outlook to future work.
\section{Background}
\label{sec:background}
In this section, we provide an overview of the relevant background on the major areas of this work: CLLMs, prompt engineering, vibe coding, agentic coding, and TDD.  
\subsection{Conversational Large Language Models}
\label{sec:CLLMs}
CLLMs synthesise or generate answers to the questions posed to them~\citep{White2023}. CLLMs can also be used for tasks such as reasoning, summarisation, translation, and coding. Current models, such as OpenAI's GPT-4~\citep{openai2024chatgpt}, typically rely on the transformer architecture~\citep{Vaswani2017} and are pre-trained~\citep{Devlin2019} on large corpora. Pre-training updates the weights in their neural network architecture through a next-token prediction technique~\citep{Ebert2023}. To use them for a specific task, the models are typically trained with task data (\ie fine-tuning), although it is not completely clear whether fine-tuning is always needed~\citep{GiagnorioEtAl2025SilverBullet}. The pre-trained models can indeed be directly interrogated (\ie prompting~\citep{Min2024}). Prompting makes use of the broad range of abilities acquired during pre-training, allowing a model to quickly adapt to a specific task at inference time, guided by natural language instructions or a few examples of the task~\citep{Brown2020}. What distinguishes CLLMs from general LLMs is their ability to maintain context across multi-turn conversations, enabling more natural and adaptive interactions. 
Since the introduction of  ChatGPT by OpenAI in November 2022, other CLLMs have been released to the general public, such as Google DeepMind, Antropic, or Meta AI; which is the most preferred is still not clear~\citep{chiang2024chatbotarenaopenplatform}. 
However, CLLMS also face limitations, including hallucinations, biases inherited from training data, and challenges in explainability.
Recent advances, such as instruction tuning, can improve their usability~\citep{White2023,White2024}. 
In this work, we use ChatGPT with an API key to be able to automatically query the CLLM from our scripts. We also use prompt templates to tune the instructions for our specific task, as described in the following. 

\subsection{Prompt engineering}
\label{promptPatterns}
A prompt is a textual instruction that guides CLLMs in generating appropriate responses. 
The effectiveness of CLLM's responses also depends on the appropriateness and informativeness of the prompt~\citep{White2023}. 
Figure~\ref{fig:promptExample} illustrates two prompts for the same code-generation task. The prompt on the left specifies only the desired functionality, whereas the prompt on the right additionally provides contextual information, a more precise input specification, and an implementation constraint, \ie the default line width.
\begin{figure}[tb]
\centering
\begin{tabular}{p{0.46\linewidth} p{0.46\linewidth}}
\textbf{Less informative prompt} &
\textbf{More informative prompt} \\[0.5em]

\fbox{\parbox[t]{\linewidth}{
Develop a class in Python called \texttt{TextFormatter} that takes arbitrary words and horizontally centers them into a line.
}}
&
\fbox{\parbox[t]{\linewidth}{
You are a professional developer in a TDD team. Develop a Python class \texttt{TextFormatter} that takes a list of words and formats them into strings of a given fixed width (default 80).
}}
\end{tabular}
\caption{Example of two prompts for the same code-generation task. The prompt on the right provides more contextual information and implementation constraints than the one on the left.}
\label{fig:promptExample}
\end{figure}
On the one hand, prompting CLLMs is fast and easy, requiring little technical expertise compared to approaches such as fine-tuning.
On the other hand, understanding the type and amount of information that a prompt should carry for the CLLM to output quality responses (\eg quality code) is a challenge. Writing prompts becomes an art.
To help design prompts, White \etal~\citep{White2023} introduced the \textit{prompt pattern}, \ie a solution to rigorously write prompts. Patterns have six components:
\begin{itemize}
    \item \textit{a name and classification.} The name provides a unique identifier for the pattern that can be referenced in discussions, and the classification groups the pattern with other patterns based on the types of problems they solve; 
    \item \textit{the intent and context.} They capture the problem that the pattern solves and the goals of the pattern; 
    \item \textit{the motivation.} It explains the rationale and importance of the problem that the pattern is solving; 
    \item \textit{the structure and key ideas.} The structure describes the fundamental contextual information that needs to be provided by the LLM to achieve the expected behaviour;
    \item \textit{example code.} It shows specific implementations of the pattern and discusses them; 
    \item \textit{consequences.} They discuss the pros and cons of using the pattern and discussion of how to adapt the pattern for different situations. 
    
\end{itemize}
In addition, a catalogue of prompt patterns has recently been proposed~\citep{White2023}.
The catalogue reports successful approaches for systematically engineering different outputs and interaction goals when working with CLLMs. The patterns and their categories are illustrated in~\tabref{tab:CLLM}. 
\begin{table}[t!b]
    \centering
    \caption{Categories and their patterns in the CLLM catalogue of \citet{White2023}.}
    \label{tab:CLLM}
    \begin{tabular}{ll}     
    \hline
\textbf{Category} & \textbf{Pattern}\\
 \hline
Input Semantics &
Meta Language Creation\\
Output Customization &
Output Automater \\&Persona\\
&Visualization Generator \\&Recipe\\
&Template\\
Error Identification &
Fact Check List \\&Reflection\\
Prompt Improvement &
Question Refinement \\&Alternative Approaches \\&Cognitive Verifier \\&Refusal\\&Breaker\\
Interaction &
Flipped Interaction Game Play\\
&Infinite Generation\\
&Context Control\\
&Context Manager\\
\hline
    \end{tabular}
\end{table}
The catalogue has also been further specialised for automating software engineering tasks (the AST catalogue, \tabref{tab:AST_pattern})~\citep{White2024}. The AST catalogue includes 14 patterns for four software engineering tasks: \textit{Requirements Elicitation, System Design and Simulation, Code Quality, and Refactoring}. No specific category and prompt patterns have been proposed for testing tasks and code generation in testing.
\begin{table}[t!b]

    \renewcommand{\arraystretch}{1.1} 
    \centering
     \caption{Classified prompt pattern for Automated Software engineering Task (AST) by \citet{White2024}.
    }
    \label{tab:AST_pattern}
    \begin{tabular}{ll}
    \hline
    \textbf{Software Engineering task}& \textbf{Pattern}\\
    \hline
        Requirements Elicitation     & Requirements Simulator \\
                                     & Specification Disambiguation \\
                                     & Change Request Simulation \\\hline
        System Design and Simulation & API Generation \\
                                     & API Simulation \\
                                     & Few-shot Example Generation \\
                                     & Domain-Specific Language (DSL) Creation \\
                                     & Architectural Possibilities \\\hline
        Code Quality                 & Code Clustering \\
                                     & Intermediate Abstraction \\
                                     & Principled Code \\
                                     & Hidden Assumptions \\\hline
        Refactoring                  & Pseudo-code Refactoring \\
                                     & Data-guided Refactoring \\\hline
    \end{tabular}
\end{table}
\subsection{Vibe coding and agentic coding}
\label{sec:agenticSystems}
Vibe coding and agentic coding follow different philosophies.
Vibe coding integrates principles from prompt
engineering, agile design, and human-CLLM co-creation to foster the collaboration between human and AI,
while abstracting much of the linguistic burden onto the
intelligent model. \textit{Vibe coding} uses intuitive, human-CLLM interaction through prompt-based, conversational
workflows. The term has been popularised by Andrej Karpathy\footnote{https://x.com/karpathy/status/1886192184808149383?lang=en} to define how developers describe what they want from a CLLM in natural language prompts, so that the CLLM generates the corresponding code.
Agentic coding, instead, aims to minimise the need for continuous human oversight. In agentic coding, AI agents autonomously initiate action, access tools and APIs, retrieve and process external data, and iteratively refine outputs through cycles of self-evaluation, making them suitable for process automation~\citep{SapkotaEtAl2025VibeAgentic}.
Agentic coding enables
autonomous software development through goal-driven AI
agents capable of planning, executing, testing, and iterating
tasks with minimal human collaboration and intervention~\citep{SapkotaEtAl2025VibeAgentic}.
Examples of tools that enable agentic coding are Google AI, Codex, Anthropic's Claude Code, whereas tools for vibe coding are the CLLMs like  OpenAI ChatGPT and Google AI Gemini. If these tools have been typically introduced as plugins for existing development environments, the latest technical advancement is that agentic online development platforms such as Replit\footnote{https://replit.com} and MetaGPT~X\footnote{\label{fn:mgx}https://mgx.dev}. 
MetaGPT~X provides multiple agents also actuated by different CLLMs. The platform can implement full-stack applications with collaborating agents that take on the roles and follow predefined workflows - called standard operating procedures (SOP) - that execute the actual development~\citep{HongEtAl2024metagpt}. 
SOPs have been introduced to avoid problems causing ``unproductive cycles'' for  multi-agent systems~\citep{GuohaoEtAl2023CAMEL}. These problems arise, for example, when cooperating independent agents engage in repetitive, contextless conversations and remain stuck while addressing complex tasks~\citep{TalebiradEtAl2023MultiAgent}.

\subsection{Test Driven Development}
\label{sec:TDD}
\begin{figure}[ht!]
    \centering
\includegraphics[width=0.9\columnwidth]{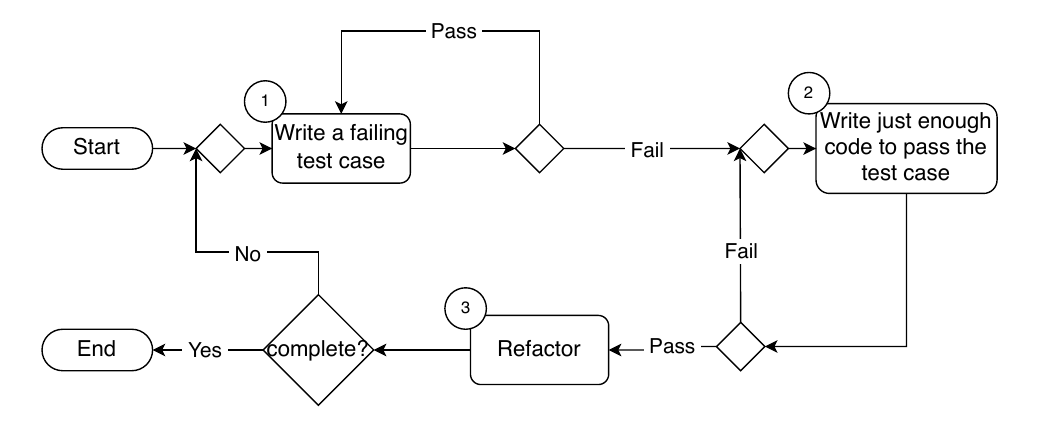}
    \caption{Overview of the TDD process.}
    \label{fig:TDD_flow}
\end{figure}

TDD is an iterative and incremental development process where tests drive the development of the business logic~\citep{Beck2022}. It consists of a sequence of micro-cycles (called iterations)~\citep{Fucci2017}, following a predefined sequence of steps: 1) writing a failing test, 2) adding the minimum production code to make the test pass, 3) refactoring the code to improve its design, as illustrated in \figref{fig:TDD_flow}. The step in which the test code is written is used to define the production code API and expected behaviour. The test is used to drive the external class interface in the intended direction.
The focus on creating the minimum code to make the current test suite pass contributes to simpler code. That, together with the safety net created by the regression tests, favours a continuous refactoring process, which has the role of continuously improving the production code's internal design.
\textit{Assertion First}~\citep{Beck2022} is a TDD pattern that suggests starting to write the test with the assertion. Each assertion should clearly define the input values, the expected output, and a pass-and-fail criterion. After defining the assertion, the rest of the test is defined to create the test scenario for that assertion. The Assertion First pattern allows the developer, when creating the test code, to initially focus on the core of what is being verified, without worrying about the test preconditions.
\section{Methodology}
\label{sec:methodology}
For any development process, we propose a four-fold methodology to: (1) design interaction models between humans and CLLMs for software development tasks; (2) define the prompts required to support these interactions; (3) model the resulting development process as a workflow; and (4) evaluate the quality of the generated artifacts and the efficiency of the development processes.
\subsection{Interaction models}\label{sec:InteractionModels}
In the following, we present the four interaction models considered in this work. Each model is described in terms of who writes the code, how the code is written, and whether the development process is automated. 
\begin{itemize}
    \item \solo: The development process is performed in the pre-AI manner, where humans write production code and test code by following development practices (\eg TDD). There is no use of CLLMs and no automation of the process;
    \item \collaborative: The human developer and the CLLM collaborate with predefined roles (\eg the human develops tests and the CLLM develops code) by following development practices (\eg TDD). This interaction model enforces vibe coding, and the automation supports the collaboration through suitable scripts;
    \item \fullyautomated: The generation process is carried out entirely by a single CLLM that writes production and test code and the process (\eg TDD) is fully automated through scripts. This model uses agentic coding, but the human designers still implement the automation of the process, and the human developers trigger the initial step of the development process via prompt; 
    \item \agentic: The development process is entirely delegated to an agentic platform.
     An agentic platform is a software framework that coordinates multiple autonomous AI agents, each assigned specialized roles and responsibilities, to collaboratively execute development tasks through predefined workflows. The development process is initially triggered by human developers, but there is no control on the agents neither on the process. 
\end{itemize}
\subsection{Prompt Design}
 \label{sec:promptDesign}
A prompt is a natural language input sequence that specifies a task for a language model, often by instantiating a template with input data and leaving one or more answer slots to be completed by the model,~\citet{LiuEtAl2023}.

We define prompts for a development task by leveraging the template catalogue proposed by \citet{White2023} and reported in \tabref{tab:CLLM}. Following \citet{LiuEtAl2023}, we construct prompts that provide sufficient contextual information and exploit the dialogic capabilities of CLLMs, without attempting to optimise them for maximum model performance, as this lies beyond the scope of this study.
Our approach is to design prompts that frame the development task as a game involving multiple players, where the objective is to generate code according to a given template. We also allow the task to be iterative so that the code can be refined. Within this setting, the CLLM may act as one or more players in the game.
Following the above approach, we can define two prompts: one for test and one for production code. To implement them, we select and combine the following relevant templates from the CLLM catalogue~\tabref{tab:CLLM}: \textit{Game Play}, \textit{Persona}, \textit{Template}, \textit{Context Manager}, and \textit{Question Refinement}.
The \textit{Game Play} together with the \textit{Persona} pattern allows us to describe the rules (\eg use \textit{Assertion First}) and the role in the TDD practice (your role is \textit{Tester}); the \textit{Template} pattern is used to layout the output (\eg provide minimal code using stubs and drivers); the \textit{Context Manager} pattern is employed to control the contextual information in which the CLLM operates (\eg ignore already implemented test cases). 
The patterns also include parameters, whose values vary depending on the stage of the development task. 
The parameters allow using the \textit{Question Refinement} pattern to iteratively update the prompt over the TDD iterations of the task, starting from the first iteration, for which not all parameters are available. 

\subsection{Workflows and their automation}
\label{sec:workflows}
We define a workflow as the operationalisation of an interaction model for a specific development process. While an interaction model identifies the actors and how they interact, a workflow specifies the activities they perform, their execution order, the interaction and decision points, the prompts and parameter values exchanged at those points, and the activities automated for execution and data collection. Applying this definition to the interaction models in \secref{sec:InteractionModels} results in four workflows:
\begin{itemize}
    \item \solo: The workflow is executed as in the pre-AI era. Only human developers execute development tasks. Any automation is  limited to collecting process logs;
    \item \collaborative: The \solo workflow is enriched through collaboration between a human and a CLLM via vibe coding. The human typically develops the tests, while the CLLM generates the production code. Assigning the tester role to the CLLM, while the human assumes the developer role, would neither relieve the human from low-level implementation details nor allow them to effectively guide and evaluate the development process. The automation is supported by prompts to query the CLLM  and scripts to collect logs and development output at different stages of the development. The human retains control over the quality of the output through monitoring;
    \item \fullyautomated: The \solo workflow is fully automated, and the generation of production and test code is delegated to the CLLM. Prompts support the automation thorough scripts that also collect logs and development output at different stages of the development. The human triggers the initial step and retains control over the quality of the output through monitoring;
    \item \agentic: The \solo workflow is delegated to an agentic platform. The human only initiates the workflow execution in the platform according to a selected configuration. The platform then plans, coordinates, and executes the development activities (processes) through its agents and its internal workflows, without human intervention. The human does not supervise or control the quality of the intermediate outputs during execution.
\end{itemize}

\begin{figure}[tb]
    \centering
    \includegraphics[width=\linewidth]{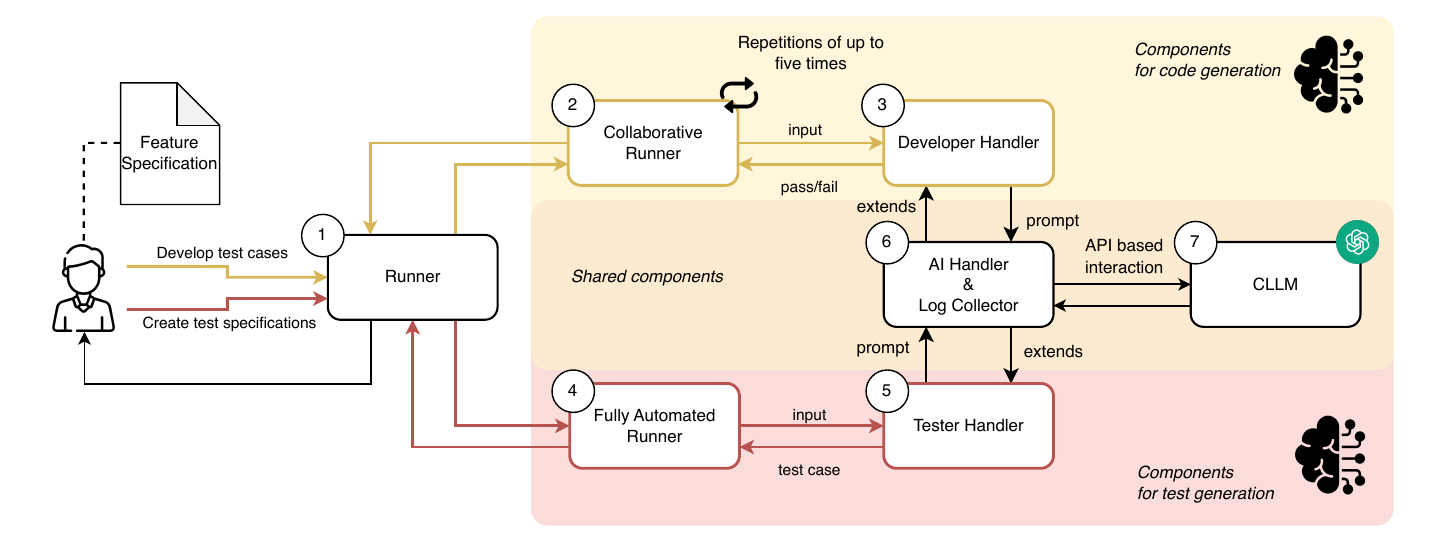}
    \caption{Overview of the toolset’s architecture.}
    \label{fig:architecture}
\end{figure}

\figref{fig:architecture} illustrates the overall automation that we designed for the  \collaborative and \fullyautomated workflows. 
In the \collaborative workflow, the human-written test code is provided to the general \textit{Runner}, step \circled{1}), which triggers the \textit{Collaborative Runner} (step \circled{2}). The \textit{Collaborative Runner} calls the \textit{Developer Handler} (step \circled{3}), which mimics the developer's activity and wraps the input (error trace, production code, and test code) into the predefined prompt. The prompt is then passed to the \textit{AI Handler} (step \circled{6}), which is responsible for the API-based interaction with the CLLM (step \circled{7}). The \textit{Collaborative Runner} updates the prompts and collects execution traces and repetition counts.

In the \fullyautomated workflow, the general \textit{Runner} triggers the \textit{Fully Automated Runner} (step \circled{4}), which calls the \textit{Tester Handler} (step \circled{5}). The \textit{Tester Handler} mimics the tester, embeds the necessary information into the predefined prompt, and passes the prompt to the shared \textit{AI Handler}, which forwards it to the CLLM (steps \circled{6} and \circled{7}). The generated test code is returned through the \textit{Fully Automated Runner} to the general \textit{Runner}, which then passes it to the \textit{Collaborative Runner} for production-code generation. Step \circled{6} also includes the shared \textit{Log Collector}, which collects the $<$error trace$>$ for both workflows.
In the \solo workflow, only logging is performed.

\subsection{Evaluation Framework}
\label{sec:evaluationFramework}
The data collected in this study comes from two complementary sources. First, we conducted a controlled pre-experimental study involving professional developers to evaluate the \solo and \collaborative workflows. Second, we replicate the study with \fullyautomated and \agentic workflows to provide exploratory evidence on autonomous software generation. To compare the output resulting from the four models of interaction,  we developed an evaluation framework  along three complementary dimensions: production code quality, test code quality, and development process efficiency. The framework defines (i) the quality attributes of interest, (ii) the metrics used to operationalize each attribute, and (iii) the statistical procedure adopted to compare the workflows.
The corresponding metrics are reported in Table~\ref{tab:mappingMeasuresUsed2RQ}.

The framework evaluates the quality of the generated production code through  complementary quality attributes. First, it assesses syntactic similarity to determine whether different workflows produce comparable implementations. It then evaluates functional correctness and structural complexity, two quality sub-attributes commonly considered when assessing generated code~\citep{TruongEtAl2026}.

The framework also evaluates the quality of the generated test suites. Specifically, it considers structural complexity and test effectiveness, measured through the coverage achieved by the generated test suites on the corresponding generated production code. These quality attributes have been identified as key indicators of generated test quality~\citep{TranEtAl2021}. Finally, the framework evaluates the efficiency of the development process through metrics describing the effort required to complete the task.

The collection of the metrics' values is performed through experiments. Thus, for each metric, we obtain four sets of values each for one workflow. We then apply statistical tests to capture difference among metrics and their related quality attributes across the four workflows.
We first apply the Kruskal-Wallis (KW) H test ~\citep{KruskalWallis1952Test} to determine whether at least one workflow differs significantly from the others. We use H as an initial screening step before performing pairwise comparisons. KW H test is a non-parametric rank-based test that checks whether different groups originate from the same distribution. The test is significant if at least one group does not come from the same distribution. The test is suitable in our settings as it does not assume a specific underlying distribution or require a particular sample size. We report the H value, the p-value, and the effect size $\epsilon^2$ for each metric to assess the significance and the magnitude of the observed differences. The effect size ranges from 0 to 1, with values of 0.01-0.08 indicating a small effect, 0.08-0.26 a medium effect, and values of at least 0.26 a large effect,~\citep{Mangiafico2025}.
To compare pairwise the four sets of values of a metric, we use one-tailed Mann-Whitney (MW) U test~\citep{MannWhitney1947Test}, reporting the U value, p-value, and Cliff's delta effect size to assess the significance and the magnitude of the observed differences. 
Cliff’s delta  ranges from -1 to 1, with 0 indicating stochastic equality of the two groups. 1 indicates that one group shows complete stochastic dominance over the other group, and a value of –1 indicates the complete stochastic domination of the other group. 
The values of the effect size 0.11-0.28 indicate a small effect, 0.28-0.43 a medium effect, and values of at least 0.43 a large effect,~\citep{Mangiafico2025}.
For each pairwise MW U test, we use $\alpha=0.05$ as the conventional significance threshold in software engineering,~\citet{KitchenhamEtAl2017}. We report the p-values and effect sizes to enable readers to interpret the results under alternative significance thresholds, in line with the recommendations of \citet{WassersteinEtAl2019Statistics}. In addition, to account for multiple pairwise comparisons on the same metric, we also apply the Holm–Bonferroni (HB) correction,~\citet{Holm1979}. Unlike the standard Bonferroni correction, HB controls the error rate of a family of tests, making it particularly suitable for studies with a limited sample size. The procedure first orders the $m$ p-values in ascending order. Starting from the smallest p-value, each is compared against the progressively less stringent significance threshold $\alpha/(m-i+1)$, where $i$ denotes its rank in the ordered list and $\alpha$ is the single-test significance threshold. The procedure stops as soon as a comparison is not significant; all subsequent hypotheses are then retained. In our study, we perform six pairwise comparisons for each metric (except in one case, where only three comparisons are possible because measurements could not be obtained for one workflow). Using $\alpha=0.05$, the corresponding HB significance thresholds are illustrated in \tabref{tab:hm}.
\begin{table}[tb]
    \centering
    \caption{HB thresholds for six and three tests with $\alpha=0.05$.}
    \label{tab:hm}
    \begin{tabular}{crr}
        \hline
        Rank&six tests &three tests\\
        \hline
     1 &  0.0083 &\\
     2 &  0.010 &\\
     3 &  0.0125& \\
     4 (1)&  0.0167 & 0.0167\\
     5 (2)&  0.0250 &0.0250\\
     6 (3)&  0.050 &0.050\\
        \hline
    \end{tabular}
\end{table}
To complement the quantitative analysis, we inspected the generated artifacts and the execution logs. We also collected feedback from the human participants to gain deeper insights into their decisions and their experience during the experiment.

\section{Empirical study}
\label{sec:experiment}

To illustrate the feasibility of our approach, we apply our methodology to the TDD process through an empirical study. We chose TDD as highly intensive and demanding development process. The major strength of TDD lies in its ability to deliver high-quality code through the granularity and uniformity of development,~\citet{Fucci2017}. To be effective, TDD developers must have a strong command of the practice and experience in development,~\citet{Causevic2011}. In this section, we describe the empirical goal and the research questions, the design of the experiments, and the automation of the TDD workflows according to the four interaction models.
\subsection{Empirical Goal and Research Questions}\label{sec:researchQuestions}
The goal of our work is to examine \textit{how CLLMs can be employed in software development}. We adopt the Goal-Question-Metric (GQM) paradigm~\citep{BasiliWeiss1984GQM} to systematically derive, from our research goal, the empirical goal (G), the research questions (Q), their associated sub-questions, and the corresponding metrics (M). The metrics are then used in the evaluation framework we introduced in \secref{sec:evaluationFramework}. 

Our main empirical goal is then: 
\par\noindent
\textit{G: Analyse  code generation  to compare human-CLLM interaction models with respect to code quality and development process efficiency, from the viewpoint of developers, in the context of TDD.}

We follow our evaluation framework in \secref{sec:evaluationFramework} to derive the RQs and the corresponding metrics as in the following.  
\par\noindent

\vspace{4pt}
\newcommand{\rqone}{\textbf{RQ1.} \textit{How does the quality of the production code created with different interaction models vary?}} 
\noindent\rqone~
To investigate this aspect, we first analyse the syntactic similarity among the production code obtained with the different workflows and then, evaluate its quality along two complementary dimensions: functional correctness, and code complexity.

\newcommand{\rqoneone}{\textbf{RQ1.1.} \textit{How does the syntactic similarity of the production code created with different interaction models vary?}}

\noindent\rqoneone~
Syntactic similarity represents the degree to which different interaction models generate independent solutions. We use the Longest Common Subsequence (\LCS) and Levenshtein Edit Distance (\LED) as in recent literature, \citep{OuyangEtAl2025NonDeterminism}.

\newcommand{\rqonetwo}{\textbf{RQ1.2.} \textit{How does the functional correctness of the production code created with different interaction models vary?}}

\noindent\rqonetwo~
Second, we focus on functional correctness, as fundamental software quality sub-attribute~\citep{ISO25010}. To evaluate correctness, we first implemented a \textit{baseline test suite} from the feature specification in \figref{fig:initialPrompt} by applying category-partition testing ~\citep{OstrandBalcer1988CPT}. Category-partition testing begins by identifying the input parameters of the function under test. The value domain of each parameter is then partitioned into categories, where each category groups values expected to trigger the same behaviour of the function. The resulting categories are then refined to eliminate redundant input combinations, and representative values are selected from each category. Using this process, we derived a set of test cases for each of the two development tasks and implemented the corresponding test methods and assertions. \listref{lst:assertions_1} and \listref{lst:assertions_2} illustrate the code for the baseline tests for the first task and the second task, respectively.
\begin{figure}[ht]
\begin{lstlisting}[
language = Python,
numbers=none,
xleftmargin=0ex,
caption={Method and its Assertions for the Task 1.},
label={lst:assertions_1}]
  def test_centerString(self):
    self.assertEqual(centerString("word", 8), "(*@{\color{codegreen}\textvisiblespace\textvisiblespace}@*)word(*@{\color{codegreen}\textvisiblespace\textvisiblespace}@*)")
    self.assertEqual(centerString("hello", 8), "(*@{\color{codegreen}\textvisiblespace}@*)hello(*@{\color{codegreen}\textvisiblespace\textvisiblespace}@*)")
    self.assertEqual(centerString("hello", 5), "hello")
    self.assertEqual(centerString("hello", 0), "")
    self.assertEqual(centerString("", 3), "(*@{\color{codegreen}\textvisiblespace\textvisiblespace\textvisiblespace}@*)")
\end{lstlisting}
\end{figure}
\begin{figure}[ht]
\begin{lstlisting}[
language = Python,
numbers=none,
xleftmargin=0ex,
caption={Method and its Assertions for Task 2.},
label={lst:assertions_2}]
  def test_centerTwoStrings(self):
    self.assertEqual(centerTwoStrings("a", "b", 5), "(*@{\color{codegreen}\textvisiblespace}@*)a(*@{\color{codegreen}\textvisiblespace}@*)b(*@{\color{codegreen}\textvisiblespace}@*)")
    self.assertEqual(centerTwoStrings("a", "b", 6), "(*@{\color{codegreen}\textvisiblespace}@*)a(*@{\color{codegreen}\textvisiblespace}@*)b(*@{\color{codegreen}\textvisiblespace\textvisiblespace}@*)")
    self.assertEqual(centerTwoStrings("a", "b", 2), "ab")
    self.assertEqual(centerTwoStrings("a", "b", 0), "")
    self.assertEqual(centerTwoStrings("", "b", 4), "(*@{\color{codegreen}\textvisiblespace\textvisiblespace}@*)b(*@{\color{codegreen}\textvisiblespace}@*)")
    self.assertEqual(centerTwoStrings("a", "", 4), "(*@{\color{codegreen}\textvisiblespace}@*)a(*@{\color{codegreen}\textvisiblespace\textvisiblespace}@*)")
    self.assertEqual(centerTwoStrings("", "", 4), "(*@{\color{codegreen}\textvisiblespace\textvisiblespace\textvisiblespace\textvisiblespace}@*)")
\end{lstlisting}
\end{figure}
We then evaluate the success rate of the baseline tests executed on the production code developed or generated in the four workflows. Specifically, we measure the Test Pass Rate (\TPR), defined as the ratio of passed baseline tests to the total number of baseline tests~\citep{OstrandBalcer1988CPT}.

\newcommand{\rqonethree}{\textbf{RQ1.3.} \textit{How does the complexity of the production code created with different interaction models vary?}}

\noindent\rqonethree~
In addition to correctness, we analyse code complexity to capture internal code characteristics that impact maintainability as sub-attribute of quality~\citep{ISO25010}. Specifically, we focus on code structural complexity and code coverage achieved by the baseline tests. To assess structural complexity, we use McCabe complexity (\MCC)~\citep{McCabe1976Complexity}, a widely used metric that quantifies decision flows in the code~\citep{Meneely2016}. For code coverage against baseline tests, we use statement coverage \SCB and branch coverage \BCB.

\par\noindent Since TDD systematically produces test code throughout the development process, evaluating the quality of the generated tests is important. We therefore compare the test suites generated in the four workflows against the baseline test suite and evaluate the coverage achieved by each generated test suite on its corresponding generated production code.

\vspace{4pt}
\newcommand{\rqtwo}{\textbf{RQ2.} \textit{How does the quality of the tests created with different interaction models vary?}}
\noindent\rqtwo~
To identify the metrics to use, we refer to the quality model defined in ~\citet{TranEtAl2021} and to the quality sub-attributes structural complexity and code coverage, which are relevant for practice,~\citet{TranEtAl2025}. 

\newcommand{\rqtwoone}{\textbf{RQ2.1.} \textit{How does the structural complexity of tests created with  different interaction models conform to that of the baseline tests?}}
\noindent\rqtwoone~
To answer this question, we use the number of test methods (\TM) and the average number of assertions per method (\TAM). 

\newcommand{\rqtwotwo}{\textbf{RQ2.2.} \textit{How does the coverage of the code created with different interaction models vary?}}
\noindent\rqtwotwo~
To answer this question, we execute the tests created with the different interaction models on the corresponding code and measure statement (\SC) and branch coverage (\BC).

\par\noindent
With the last research question, we shift our focus from the product to the process.

\vspace{4pt}
\newcommand{\rqthree}{\textbf{RQ3.} \textit{How does the efficiency of the TDD process vary between different interaction models?}}

\noindent\rqthree~
We measure process efficiency as the effort required to complete a task, measured through the number of iterations (\mIterations) and the completion time, in line with definitions of efficiency as performance relative to resource usage~\citep{ISO25010} and studies on performance in solving program-comprehension tasks~\citep{AstromskisEtAl2017,JanetEtAl2014}.
Specifically, we employ the number of TDD iterations (\mIterations) and completion time (\Time), which is both self-reported by participants in the human-centred models and collected from the runtime in the automated models.

\tabref{tab:mappingMeasuresUsed2RQ} summarises the mapping between the three research questions and the  metrics used to answer them and define our evaluation framework described in \secref{sec:evaluationFramework}. Each metric is presented with its acronym, name and description. The table also reports CLOC and TLOC for completeness and to provide an indication of the size of the generated artifacts.

\begin{table}[t!b]
    \renewcommand{\arraystretch}{1.1} 
    \setlength{\tabcolsep}{6pt} 
    \newcolumntype{P}[1]{>{\raggedright\arraybackslash}p{#1}}
    \centering
    \caption{Metrics per research question.}
    \label{tab:mappingMeasuresUsed2RQ}
    \begin{tabular}{p{0.05\linewidth}p{0.1\linewidth}P{0.15\linewidth}p{0.55\linewidth}} \hline
        RQ&Acronym& Name & Definition\\\hline
        & CLOC &  Code LOC & Number of lines in the production code, empty lines and comments were not counted, positive integer.\\
        & TLOC &  Test LOC & Number of lines in the test code, empty lines and comments were not counted, positive integer.\\
        \hline
        RQ1 & \LCS & Longest Common Subsequence & Length of the longest common subsequence between two sequences;  positive integer. \\
         & \LED & Levenshtein Edit Distance & Minimum number of single-token edits (insertions, deletions, or substitutions) required to change one text into the another; positive integer. \citep{Levenshtein1966}.\\
         & \TPR & Test Pass Rate & Proportion of baseline test cases that pass on the production code over the total number of baseline tests; percentage from 0 to 100.\\
         & \MCC & McCabe Complexity & Number of code branches plus one; an integer greater than one. \citep{McCabe1976Complexity} \\
         & \SCB & Statement coverage with respect to baseline tests& Proportion of executable statements in the production code that are covered by the baseline tests \citep{ZhuEtAl1997Coverage};  percentage from 0 to 100.\\
         & \BCB & Branch coverage with respect to baseline tests& Proportion of executable branches in the production code covered by the baseline tests \citep{ZhuEtAl1997Coverage}; percentage from 0 to 100.\\\hline
        RQ2 & \TAM & Test assertions per method & Average number of assertions per test method; positive real number.\\       
         & \TM & Number of test methods & Number of test methods;  positive integer.\\
         & \SC & Statement coverage of the created tests& Proportion of executable statements in the production code covered by tests created in the workflow; \citep{ZhuEtAl1997Coverage}; percentage from 0 to 100.\\
         & \BC & Branch coverage of the created tests& Proportion of executable branches in the production code covered by tests created in the workflow; \citep{ZhuEtAl1997Coverage}; percentage from 0 to 100.\\\hline
        RQ3 & \mIterations & Number of iterations & Number of TDD iterations recorded by the toolset during the experiment; positive integer.\\
         & \Time & Time & Time required to complete the task (in seconds), self-reported by participants in \solo and \collaborative workflows, and collected from the runtime in \fullyautomated and \agentic workflows; positive integer. \\
        \hline
    \end{tabular}
\end{table}
\subsection{Design of the experiment}
\label{sec:designExperiment}
Firstly, we conducted a \textit{preliminary experiment} with five professional developers~\citep{MockEtal2024TDD} to assess the feasibility of the study and collect feedback on the experimental protocol. Based on the insights gained from the preliminary experiment, for the study reported in this paper, we extended the experiment with an additional 11 professionals, using the same toolset and experimental procedure and task. 
Although the total sample size falls within the typical range for empirical studies involving professional software engineers (15–25 participants~\citep{KitchenhamMadeyski2024}), the results remain subject to the limitations associated with a relatively small sample of human participants. To mitigate potential threats arising from this limitation, we adopted a threefold strategy. First, we followed a pre-experimental design ~\citep{CampbellEtAl2015}. 
Second, we employed non-parametric statistical tests and controlled statistical power through multi test significance with significance threshold correction (\secref{sec:evaluationFramework}).
Third, we complemented the statistical analysis with a careful inspection of the data to account for potential small-sample effects. Nevertheless, we acknowledge that the limited sample size may reduce the statistical power of the comparisons involving the human participants. We discuss this limitation and its implications in the Threats to Validity section (\secref{sec:threatValidity}).
We executed the \fullyautomated and \agentic workflows on the same development tasks. 
For the collaborative and fully automated workflows, we used OpenAI’s \textit{GPT-3.5 Turbo} through its API.

In all cases, we selected the same two development tasks that did not require  understanding complex APIs or class structures. \figref{fig:initialPrompt} reports them.
\begin{figure}[ht!]
    \centering
   \begin{mdframed}[backgroundcolor=gray!5, nobreak=true]\
The goal of this experiment is to develop in Python the following feature:
\begin{center}
\textit{Develop a class TextFormatter that takes arbitrary words and horizontally centres them into a line.}
\end{center}
\par\noindent
The class TextFormatter shall have three functions:
\begin{itemize}
\item[-] setLineWidth sets the length of the line. 
\item[-]
The second function receives a single word and returns the word in the centre of the line. \textit{First task.}
\item[-]  The third function receives two words and centres the two words in the line. \textit{Second task.}
\end{itemize}
\par\noindent
To develop it you will use \textit{Test Driven Development with assertion first}.
\end{mdframed}
    \caption{Tasks' specification.}
    \label{fig:initialPrompt}
\end{figure}

\subsubsection{Experiment with professionals.}
\label{sec:study_context} 
We applied the Static-Group Comparison, as pre-experimental design with control group~\citep{CampbellEtAl2015,Wohlin2012}, with the purpose of establishing the effect of a treatment X, \figref{fig:pre_experiment}. 
Treatment \textit{X} in our experiment is the application of the \collaborative workflow to the TDD process. 
The design then foresees two groups: the Experimental Group (\textit{EG}) and the Control Group (\textit{CG}), where the EG follows  \collaborative model and the CG the  \solo one.
Human participants in the \collaborative and \solo groups are randomly assigned and given the same tasks. 
The metrics  are collected only once during the experiment and analysed after the experiment (\textit{O}) to determine the effects of treatment (\textit{X}).
\begin{figure}[!ht]
\centering
\caption{The Static-Group Comparison with control group: \textit{EG} = experimental group and \textit{CG} = control group, \textit{X} = treatment, and \textit{O} = ex-post observation.}
\label{fig:pre_experiment}
\begin{tabular}{lccc}\\
\textit{EG } &&X&O  \\
\cdashline{2-4}
\textit{CG } &&&O 
\end{tabular}
\end{figure}
At the end of the experiment, participants also completed a post-experiment questionnaire to assess the perceived difficulty of the task and their overall experience with the development process.  %
\paragraph*{Participants.}~In the experiments, we only employed IT professionals. 
Employing professionals in software engineering experiments presents both advantages and drawbacks~\citep{RomanoEtAl2025,SalmanEtAl2015}. On the positive side, their participation yields more realistic results, as their behaviour more closely mirrors that of developers in industry.
Their experience helps make results more connected to the actual practice. On the negative side, their behaviour is less predictable because they have different backgrounds and habits, so results may be harder to control. It is also difficult to get them involved, as recruiting them requires more time and effort, and their busy schedules often limit their participation~\citep{RomanoEtAl2025}. 
The typical sample size of studies involving professionals ranges from 15 to 25 participants~\citep{KitchenhamMadeyski2024}. 
With 16 professionals, our study falls well within this range. The participants were IT professionals from Italy, Denmark, and Brazil, with experience in TDD and Python, as illustrated in \tabref{tab:participantsDemographics}. 
All participants were required to complete the same development tasks. They were introduced to the experiment by one of the authors. 
\begin{table}[t!b]
    \renewcommand{\arraystretch}{1.1} 
    \centering
    \caption{Demographics of the participants.}
    \label{tab:participantsDemographics}
    \begin{tabular}{cccl}
    \hline
    ID & TDD experience & Python experience & Role \\\hline
    \multicolumn{4}{c}{\cellcolor{gray!15}\solo participants} \\\hline
    P1 & 1-3 years   & $>$3 years & Software developer \\
    P2 & 1-3 years   & $<$1 year  & Software developer \\
    P3 & $<$1 year   & $>$3 years & Security software engineer\\
    P4 & $<$1 year   & $<$1 year  & DevOps engineer    \\
    P5 & 1-3 years   & $>$3 years & Software developer \\
    P6 & $<$1 year   & $<$1 year  & Software developer \\
    P7 & 1-3 years   & $>$3 years & DevOps engineer    \\
    P8 & $<$1 year   & $<$1 year  & Software developer \\
    P9 & $<$1 year   & 1-3 years  & Software developer \\\hline
    \multicolumn{4}{c}{\cellcolor{gray!15}\collaborative participants} \\\hline
    P10 & $>$3 years  & $<$1 year  & Software developer \\
    P11 & 1-3 years   & $>$3 years & Data scientist     \\
    P12 & $<$1 year   & $<$1 year  & Software engineering \\
    P13 & $<$1 year   & $<$1 year  & Software developer \\
    P14 & $<$1 year   & $<$1 year  & Technical leader   \\
    P15 & $>$3 years  & $>$3 years & Software developer \\
    P16 & $<$1 year   & 1-3 years  & Software developer \\
    \hline
    \end{tabular}
\end{table}
\paragraph*{Study Protocol.}
~We carried out the experiment in two ways: \textit{remote}, using a call and screen sharing and \textit{in-person}. 
In both cases, we used Google Colab\footnote{\label{fn:colab}\href{https://colab.research.google.com/}{https://colab.research.google.com/}} as the development environment, \ie a cloud-based service that provides computing resources through a Jupyter notebook interface. Using this tool, we were able to control dependencies, ensure the same Python interpreter version, and create a script to collect data during the experiment.
Before the experiment, we obtained the consent of the participants: they were informed of their voluntary participation and their right to withdraw at any time without providing a reason. The participants were introduced to the tasks and Google Colab. 
During the experiment, participants were allowed to search the Internet. However, they were prohibited from searching for complete solutions or independently querying any CLLM, \eg ChatGPT. This restriction was enforced by monitoring participants via screen sharing or direct observation in the in-person setting. Although this allowed us to verify that participants did not use external AI tools, we cannot completely rule out the possibility that someone may have used an AI tool, \eg on a second screen. 
The participants were given 40 minutes to complete the tasks. During the experiment, one of the authors was available to assist the participants with environmental issues or task clarification. No assistance was provided on how to solve the tasks. Participants who interact with the CLLM were not allowed to modify the prompts. 
After completing the tasks, participants were asked to submit their code and complete a feedback questionnaire, which can be found in the replication package \citep{MockEtAl2025journal}.
\subsubsection{Experiment with agents}
\label{sec:experimentsWithAgents}

For the \agentic workflow, we selected MetaGPT~X (MGX\footnote{MGX was used through the web interface\textsuperscript{\ref{fn:mgx}}.}) as a representative for agentic platforms. The platform has two operation modes: Engineer and Team. To resemble our experiment the most, we chose the Engineer mode (single developer).
In repeated individual runs, we configured the platform to use one of two models: OpenAI ChatGPT 5\footnote{https://openai.com/index/introducing-gpt-5/} and Anthropic Claude Sonnet 4\footnote{https://www.anthropic.com/news/claude-4}.
ChatGPT 5 is the most similar to the model we used in the experiment, while Claude Sonnet 4 is one of the most promising and suitable models in our context~\citep{PallaSlaby2025,NascimentoEtAl2025}.
We collected the generated code and computed the same metrics as for the participants. However, we could not collect the number of iterations because the MetaGPT~X operates with standard operating procedures (SOPs), which do not support a TDD-like interaction with the user; therefore, the solution was provided in a single response. For the \agentic workflow we have defined two prompt scenarios: 1) we provided a minimal prompt and 2) we provided a more detailed description of the prompt, like the one used for the \fullyautomated workflow. 

In \tabref{tab:agentic_category_means}, the mean value of each metric (see \tabref{tab:mappingMeasuresUsed2RQ}) is computed for each model and prompt scenario. We observe little difference between the two CLLMs. However, between the two prompt scenarios, minor variations can be observed for some metrics, \eg \MCC, \TAM, and \TM. Additionally, \CLOC and \TLOC vary across the prompt scenarios. Overall, these observations mainly relate to the structure of the generated code rather than its functionality. Interestingly, most coverage metrics (\SCB, \BCB, and \SC) remain similar across the four different settings, suggesting that neither the choice of model nor the prompting scenario substantially affects the overall testability of the generated code or the effectiveness of the generated test suites. Given the minor differences in \MCC, \TAM, and \TM, we decided to merge all \agentic workflow executions and analyse them together rather than individually. Future work is needed to investigate the performance differences of the \agentic workflow under different prompting settings. Therefore, this part of the study should be considered exploratory.
\begin{table}[ht]
    \renewcommand{\arraystretch}{1.1} 
    \centering
    \footnotesize
    \setlength{\tabcolsep}{2pt}
    \renewcommand{\arraystretch}{0.95}
    \caption{Mean metric values for the \agentic runs, grouped by model and prompt scenario. CG = ChatGPT 5, CS = Claude Sonnet 4.}
    \label{tab:agentic_category_means}
    \begin{tabular}{C{0.075\textwidth}|C{0.075\textwidth}C{0.06\textwidth}cC{0.06\textwidth}C{0.06\textwidth}|C{0.075\textwidth}C{0.10\textwidth}cC{0.06\textwidth}C{0.06\textwidth}|cC{0.065\textwidth}}
    \hline
    ~ & \multicolumn{5}{c|}{Production code} & \multicolumn{5}{c|}{Test code} & \multicolumn{2}{c}{Process}\\
    \hline
     & \CLOC & \TPR ($\%$) & \MCC & \SCB (\%) & \BCB (\%) & \TLOC & \TAM(Ass.) & \TM & \SC (\%) & \BC (\%) & Iter. & \Time (sec) \\\hline
    CG1 & 53.67 & 94.33 & 15.00 & 63.33 & 60.67 & 41.33 & 7.48  & 3.67 & 97.00 & 92.33 & - & 120 \\
    CG2 & 15.67 & 92.00 & 9.67 & 61.00 & 67.33 & 22.33 & 1.00  & 3.00 & 85.67 & 50.00 & - & 60 \\
    CS1 & 53.00 & 97.33 & 13.67 & 64.33 & 67.00 & 35.33 & 5.77  & 7.00 & 83.33 & 75.33 & - & 60 \\
    CS2 & 21.00 & 94.67 & 8.33 & 71.00 & 66.33 & 25.33 & 2.00  & 4.33 & 85.67 & 50.00 & - & 60 \\\hline
    \end{tabular}
\end{table}

For the \fullyautomated workflow we created an automated process as described in \secref{sec:automatingTDD} and \figref{fig:fullyAutomated}. The workflow executed in a CLI setting, instructed with OpenAI’s \textit{GPT-3.5 Turbo} through its API. It was configured with the prompts listed in \tabref{tab:prompts}; the prompts and their iterative refinements are embedded in the toolset. During execution, the workflow automatically recorded all logs, including the TDD iterations and their intermediate solutions. Execution time was measured using the terminal command \texttt{time}, which was prefixed to the CLI execution command. All experiments were conducted on a macOS host machine. As with the \agentic workflow, we collected the generated code and computed the same evaluation metrics as those used for the participant submissions. The workflow was executed multiple times with the same settings.
\subsection{Implementing the TDD workflows}
\label{sec:automatingTDD}
\begin{figure*}[b]
    \centering
    \includegraphics[width=\textwidth]{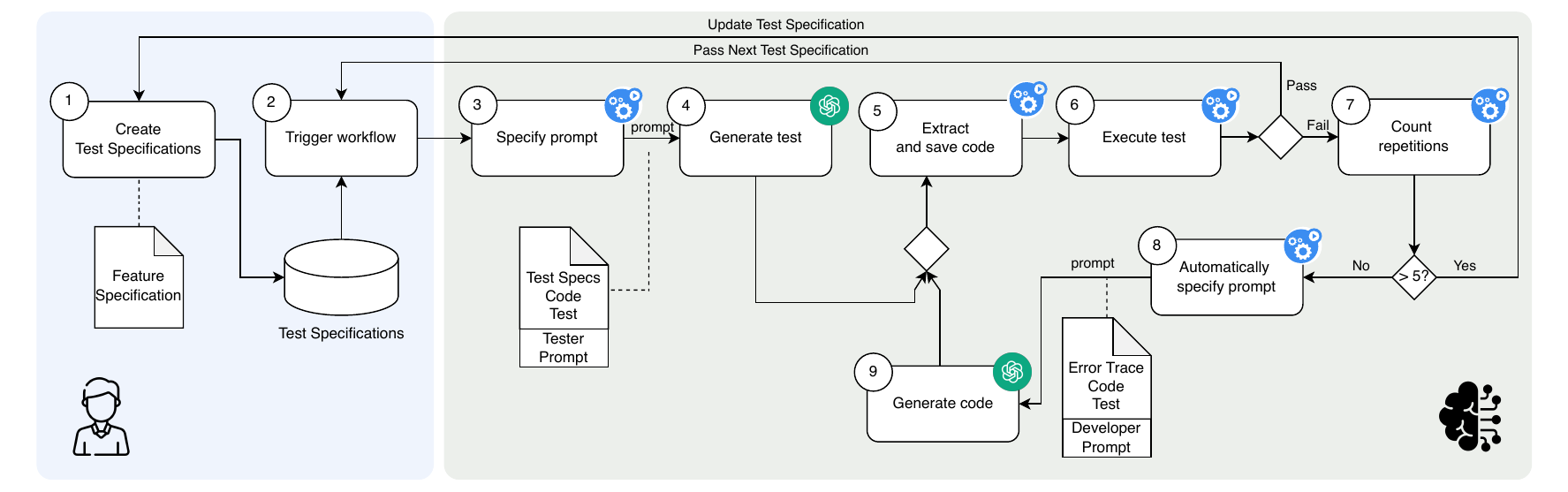}
    \caption{\fullyautomated workflow.}
    \label{fig:fullyAutomated}
\end{figure*}

\begin{figure*}[tb]
    \centering
    \begin{minipage}[t]{0.66\textwidth}
        \centering
        \includegraphics[width=\linewidth]{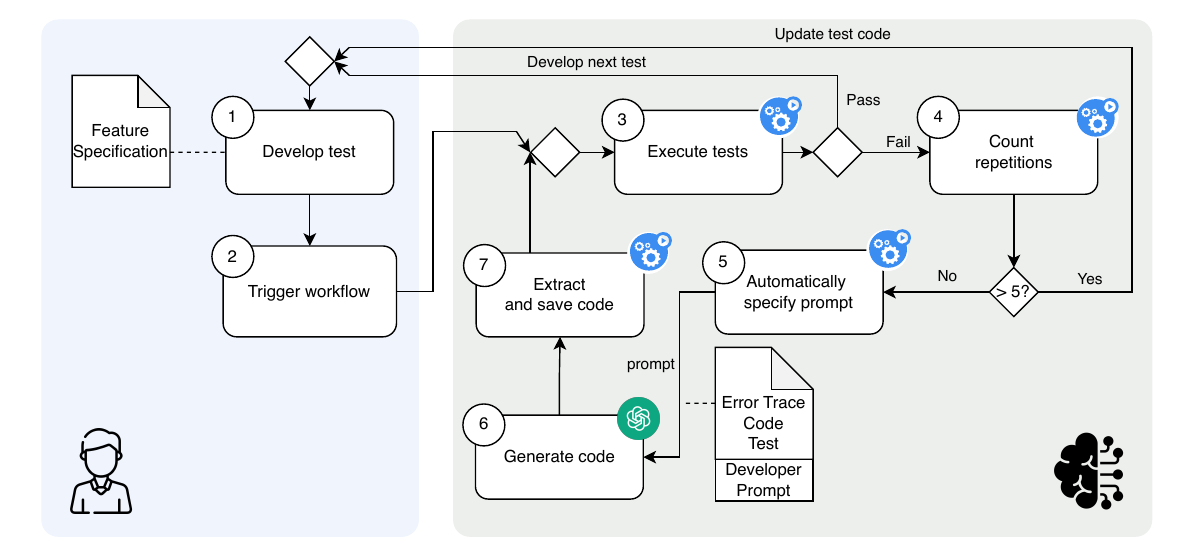}
        \caption{\collaborative workflow.}
        \label{fig:collaborative}
    \end{minipage}
    \hspace{0pt}
    \raisebox{-3pt}{%
    \begin{minipage}[t]{0.317\textwidth}
        \centering
        \includegraphics[width=\linewidth]{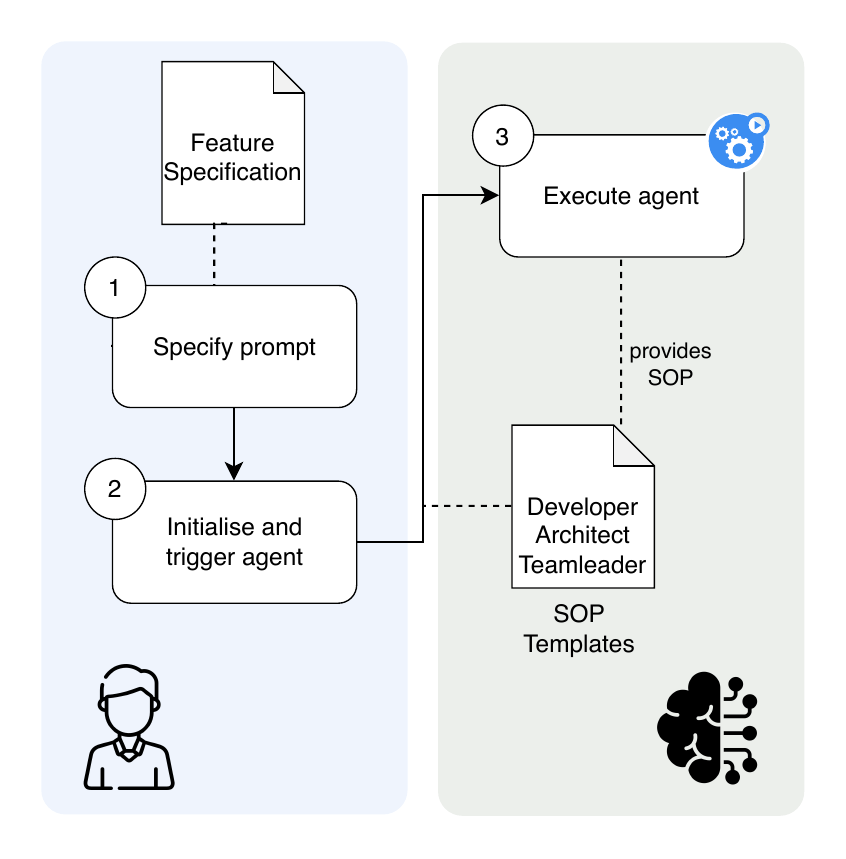}
        \vspace{-13pt}
        \caption{\agentic workflow.}
        \label{fig:agentic}
    \end{minipage}%
    }
\end{figure*}

In this section, we define two TDD workflows (\figref{fig:fullyAutomated}, \figref{fig:collaborative}), corresponding to the \fullyautomated and \collaborative interaction models.
In the figures, activities marked with the OpenAI symbol are performed by the CLLM, while those with the gear symbol are executed by our scripts; all remaining activities are carried out by humans. The note symbols indicate the prompts used to query the model, as described in \tabref{tab:prompts}.
In the workflows, an iteration is the execution of the workflow from the creation of a single assertion until the generated code passes it. A repetition, instead, refers to a single loop cycle in which the workflow is executed for a specific assertion to verify whether the generated code satisfies it.
The two interaction workflows only differ in the way the tests are created. The \fullyautomated workflow is initiated by humans with a set of test specifications, \circled{1}\footnote{It is worth noting that we do not automate the generation of the test specifications or any other natural language artefact, as we want to keep the focus on the generation of code.}. The human then selects one test specification and triggers the workflow \circled{2}, one for the first iteration and the other for subsequent iterations, as in \tabref{tab:prompts}, \circled{3}. Afterwards, the CLLM is prompted for the generation of the test code \circled{4}. Then, our toolset automatically executes the test against any available production code and collects its execution traces, \circled{5} and \circled{6}.

From this point onward, the two TDD workflows of the two interaction models (\fullyautomated and \collaborative) perform the same activities in a loop, where the CLLM acts as a developer and generates the production code.
Whenever the tests fail – for example, during the first iteration when no production code exists yet - the loop (activities \circled{7} to \circled{9} in \figref{fig:fullyAutomated} and \circled{4} to \circled{7} in \figref{fig:collaborative}) is triggered. The same prompt is automatically reissued up to five times within the same interaction session, using modified prompts from the second iteration onwards~\tabref{tab:prompts}. If the CLLM is still unable to generate production code that satisfies the test code, the loop is interrupted and execution is redirected. In this case, in the \collaborative mode, where humans act as testers, we allow them to refactor the generated code at their convenience, following TDD. In the \fullyautomated model, we let the CLLM handle the refactoring autonomously by simply instructing it to follow the TDD practice.
From the second iteration onward, the entire existing test and production code is passed into the prompts \circled{4} and \circled{9}, and \circled{8} in ~\figref{fig:fullyAutomated} and ~\figref{fig:collaborative}, respectively. 

\begin{figure*}[!ht]
    \centering
    \includegraphics[width=\textwidth,trim={0 9cm 0 9cm},clip]{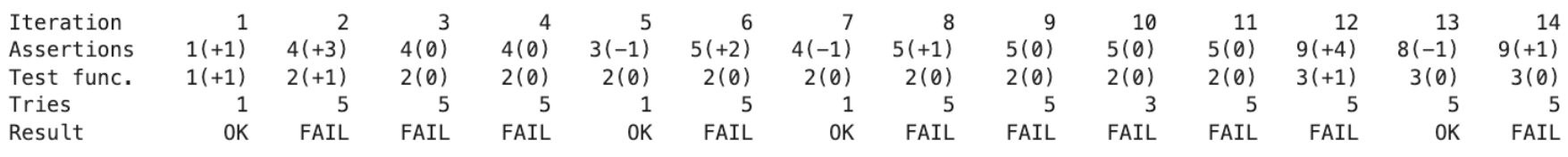}
    \caption{Example of log data in \collaborative process.}
    \label{fig:logs}
\end{figure*}

The execution data we obtained is stored in logs. 
\figref{fig:logs} shows the log for participant P15 with \collaborative workflow and 14 iterations (columns). 
The first row describes the cumulative number of assertions developed by the participant. The second row reports the cumulative number of test functions. The third row counts the repetitions and the last row reports the output of the test execution. 
In the first iteration, the CLLM generates code that passes the test method developed by the participant (``OK''). In the second iteration,  the participant develops a new test function, with three assertions. For this test case, the CLLM was unable to generate production code that passes the tests in five attempts. Thus, the participant iterated the generation for two further times (iterations 3 and 4), before the participant modified the test cases in iteration 5.  

\subsubsection{Prompts}\label{TDDPrompts}
\tabref{tab:prompts} illustrates the prompts we designed for the workflows \collaborative and \fullyautomated.  
Colours identify the relevant templates as listed in \tabref{tab:CLLM}, while angular brackets $<$\dots$>$ identify the parameters.
For the \collaborative and \fullyautomated models, the table also shows two types of prompts: the one used in the first iteration and the one used in subsequent iterations of the TDD process. 

\begingroup
    \renewcommand{\arraystretch}{1.1} 
    \setlength{\LTcapwidth}{\textwidth}
\begin{longtable}{>{\footnotesize\arraybackslash}p{\dimexpr\textwidth-2\tabcolsep\relax}}
    \caption{Prompts designed for the \collaborative, \fullyautomated, and \agentic interaction workflows of the TDD process.}
    \label{tab:prompts}\\
    \hline

\hspace*{\fill}\textbf{Collaborative}\hspace*{\fill} 
    \vspace{1pt}
    
    The human derives the test case specifications from the feature. At each iteration, the human develops a test code - \ie one assertion and the surrounding test code that ensures its execution - for a test case specification. At the end of each iteration, the test code is executed and a new error trace is produced.
    \vspace{4pt}
      
    \textbf{First iteration};  \textbf{CLLM role: developer}\\
    \textcolor{Orange}{You are part of a TDD team.} \textcolor{Green}{Your role is  developer.} Based on the information provided below, write the \textcolor{Brown}{minimal production code necessary to ensure that the test case no longer fails.} \textcolor{RoyalBlue}{Built-in functions are not allowed.}
     
\hspace*{\fill}$<$error trace, test code$>$
\hspace*{\fill}
\vspace{2pt}
    
\textbf{Second and further iterations}; \textbf{CLLM role: developer} 

\textcolor{Orange}{You are part of a TDD team.} \textcolor{Green}{Your role is  developer}. Based on the information provided below, write the \textcolor{Brown}{minimal production code necessary to ensure that the test case no longer fails.} \textcolor{RoyalBlue}{Built-in functions are not allowed.}
\vspace{2pt}

\hspace*{\fill}
   $<$error trace, test code, \textcolor{magenta}{existing test code, existing production code}$>$
\hspace*{\fill}
\vspace{2pt}
\\\hline

\hspace*{\fill}\textbf{Fully automated} \hspace*{\fill}
\vspace{1pt}

At each iteration, the CLLM is given one baseline test case specification. The CLLM develops the test and the production code. At the end of each iteration, the test code is executed, and a new error trace is produced. 
\vspace{4pt}

\textbf{First iteration};  \textbf{CLLM role: tester}

\textcolor{Orange}{You are part of a TDD team}. \textcolor{Green}{Your role is  tester}. Based on the information provided below, write  \textcolor{Brown}{test code using the `unittest` library}.
\textcolor{Orange}{Use Assertion-First in TDD.}
\textcolor{RoyalBlue}{
Focus only on the test case specification described below. Do not implement solutions for other test cases.}
\vspace{2pt}

\hspace*{\fill} $<$test case specification$>$ \hspace*{\fill}
\vspace{2pt}

\textbf{CLLM role: developer}

\textcolor{Orange}{You are part of a TDD team.} \textcolor{Green}{Your role is  developer}. Based on the information provided below, write the \textcolor{Brown}{minimal production code necessary to ensure that the test case no longer fails}. \textcolor{RoyalBlue}{The code must be written in the same file of the test.} \textcolor{RoyalBlue}{Built-in functions are not allowed.}
\vspace{2pt}

\hspace*{\fill}$<$test code, error trace$>$\hspace*{\fill}
\vspace{2pt}

\textbf{Second and further iterations}; \textbf{CLLM role: tester} 

\textcolor{Orange}{You are part of a TDD team.} \textcolor{Green}{Your role is  tester}. Based on the information provided below, \textcolor{Brown}{write test code using  `unittest` library}. 
 \textcolor{Orange}{Use  Assertion-First  in TDD.} 
\textcolor{RoyalBlue}{Keep all existing tests intact and add one new test case.  Focus only on the
test case specification described below. Do not implement solutions for other test cases.}
\vspace{2pt}

\hspace*{\fill} $<$test case specification, \textcolor{magenta}{existing production code, existing test code}$>$\hspace*{\fill}
\vspace{2pt}

\textbf{CLLM role: developer}

\textcolor{Orange}{You are part of a TDD team}. \textcolor{Green}{Your role is  developer}. Based on the information provided below, write the \textcolor{Brown}{minimal production code necessary to ensure that the test case no longer fails}. 
\textcolor{RoyalBlue}{The code must be written in the same file of the test.} \textcolor{RoyalBlue}{Built-in functions are not allowed.}
\vspace{2pt}

\hspace*{\fill}$<$test code, error trace,  \textcolor{magenta}{existing test code, existing production code}$>$\hspace*{\fill}
\vspace{2pt}
\\\hline

\hspace*{\fill}\textbf{Agentic} \hspace*{\fill}
\vspace{1pt}

The agent is provided with an initial prompt. We experimented with two different prompts. 
\vspace{4pt}

\textbf{Prompt 1} 

Develop the following feature: 
\vspace{2pt}

\hspace*{\fill} $<$task specification$>$\hspace*{\fill} 
\vspace{2pt}

To develop it you will use \textcolor{Orange}{TDD with assertion first}.
\vspace{2pt}

\textbf{Prompt 2}

\textcolor{Orange}{You are part of a TDD team}.
Based on the information provided below, \textcolor{Brown}{write test code using  `unittest` library}.
\textcolor{Orange}{Use Assertion-First in TDD}
\textcolor{RoyalBlue}{The code must be written in the same file of the test.  Focus only on the
test case specification described below. Do not implement solutions for other test cases.}
\vspace{2pt}

\hspace*{\fill} $<$task specification$>$\hspace*{\fill} \\\hline

 \textbf{Templates:} \textcolor{Orange}{Game Play}, \textcolor{Green}{Persona}, \textcolor{Brown}{Template}, \textcolor{RoyalBlue}{Context Manager}, \textcolor{magenta}{Question Refinement}\vspace{2pt}
\\\hline
\end{longtable}
\endgroup
For example, in the first iteration of the \collaborative workflow, no production code is yet available; therefore, the prompt parameters consist only of $<$test code$>$ and $<$error trace$>$. 

\begin{figure}[ht]
\begin{lstlisting}[
    caption={Example of test code and error trace with one assertion for the feature that develops the class TextFormatter.}, 
    label={lst:input}
]
# Test code: 
import unittest
class TestTextFormatter(unittest.TestCase):
    def test_center_word(self):
        f = TextFormatter()
        f.setLineWidth(10) 
        centered_string = f.center('word') 
        self.assertEqual(centered_string,'(*@{\color{codegreen}\textvisiblespace\textvisiblespace\textvisiblespace}@*)word(*@{\color{codegreen}\textvisiblespace\textvisiblespace\textvisiblespace}@*)') 
if __name__ == '__main__':
    unittest.main()
# Error Trace
Traceback (most recent call last):
  File "$<$path$>$", line 18, in test_center_word
    self.assertEqual(centered_string,'(*@{\color{codegreen}\textvisiblespace\textvisiblespace\textvisiblespace}@*)word(*@{\color{codegreen}\textvisiblespace\textvisiblespace\textvisiblespace}@*)') 
AssertionError: None != '(*@{\color{codegreen}\textvisiblespace\textvisiblespace\textvisiblespace}@*)word(*@{\color{codegreen}\textvisiblespace\textvisiblespace\textvisiblespace}@*)'
\end{lstlisting}
\end{figure}
\begin{figure}[ht]
\begin{lstlisting}[
    caption={Example of generated code with the free online version of ChatGPT 3.5-turbo.}, label={lst:code}
]
class TextFormatter:
    def setLineWidth(self, width):
        pass
    def center(self, string):
        if len(string) $>$= 10:
            return string
        else:
            num_spaces = (10 - len(string)) // 2
            return ' ' * num_spaces + string + ' ' * num_spaces
\end{lstlisting}
\end{figure}
An example of such input is shown in \listref{lst:input}. 
From the CLLM’s response, the production code is extracted; an example is provided in \listref{lst:code}.

In subsequent iterations, the production code is included in the prompt (see \tabref{tab:prompts}) together with the $<$test code$>$ and $<$error trace$>$. Additionally, prompts incorporate the production and test code generated in previous iterations. This enables the incremental generation of both code and tests throughout the process. Each iteration is executed as a new interaction session to prevent the CLLM from being biased by prior iterations.

For the \fullyautomated model, at each iteration, the CLLM is given in sequence one of the following test specifications:
\vspace{2pt}
\begin{itemize}
    \item line width=10 and word=``word",
    \item line width=10 and word=``hello",
    \item line width=10 and words=[``foo", ``bar"]
\end{itemize}
For the \agentic workflow, \tabref{tab:prompts} illustrates the initial prompts each triggering the entire development task. No other prompts have been used. 

\section{Results}
\label{sec:experimentResults}
From the RQs, we derived the metrics of our evaluation framework~\tabref{tab:mappingMeasuresUsed2RQ}. We then applied the statistical analysis defined by the framework, beginning with the H test on all metrics. \tabref{tab:KruskalWallis} shows that 
H is statistically significant for CLOC, \TPR, \MCC, \BCB, \SC, \BC, and \Time, with large effect size  for CLOC, \TPR, \MCC, and \Time and medium effect size for \BCB, \SC, and \BC.
\begin{table}[t!b]
    \renewcommand{\arraystretch}{1.1} 
    \centering
    \caption{Kruskal-Wallis H test across models. The asterisk (*) denote statistical significance at $\alpha=0.05$; effect size computed with $\epsilon ^2$. 
    The comparison for \mIterations excludes the \agentic workflow.}
    \label{tab:KruskalWallis}
    \begin{tabular}{llccl}
    \hline
            & Metric                    &         H  &  Effect Size &    p-value \\\hline
        & CLOC                  &      12.27 &         0.34 &       \num{6.52e-03}* \\
        & TLOC                  &       6.22 &         0.12 &         \num{1.01e-01} \\ \hline
       RQ1 & \TPR               &      16.98 &         0.52 &        \num{7.12e-04}* \\
       & \MCC                       &      16.59 &          0.5 &        \num{8.58e-04}* \\
        & \SCB    &       4.88 &         0.07 &       \num{1.81e-01}\\
        & \BCB      &       9.37 &         0.24 &       \num{2.48e-02}* \\\hline
       RQ2 & \TAM               &       6.44 &         0.13 &       \num{9.22e-02} \\
        & \TM         &       1.86 &        -0.04 &        \num{6.02e-01} \\
        &\SC   &       7.99 &         0.18 &       \num{4.62e-02}*\\
        &\BC      &        8.0 &         0.19 &      \num{4.60e-02}*\\\hline
        RQ3 & \mIterations                &       4.24 &         0.14 &       \num{1.20e-01} \\
        & \Time                       &      20.13 &         0.63 &        \num{1.60e-04}*\\\hline
    \end{tabular}
\end{table}
The test therefore indicates that, for at least one of the workflows, the production code, test code, or development process differs with respect to the attributes measured by the selected metrics.
For example, \figref{fig:boxplotsLOC} illustrates why the null hypothesis is rejected for CLOC but not for TLOC. The boxplot of  CLOC for the \agentic workflow is clearly separated from those of the other workflows, whereas the boxplots of TLOC overlap to varying degrees across all workflows.
\begin{figure}[tb]
    \renewcommand{\arraystretch}{1.1} 
    \centering
    \includegraphics[width=\linewidth]{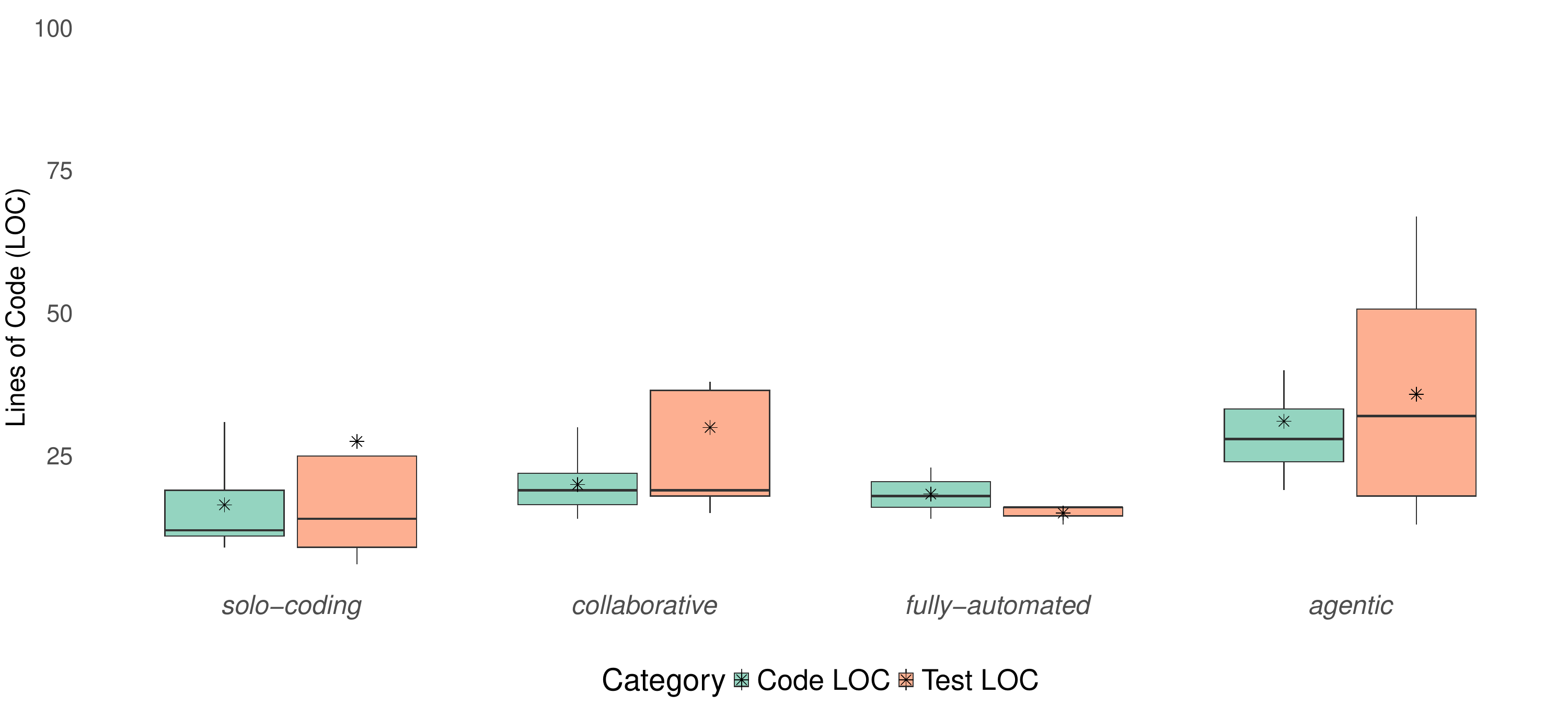}
    \caption{Boxplots of code and TLOC grouped by workflow. The asterisk (*) denotes the mean value.}
    \label{fig:boxplotsLOC}
\end{figure}
\subsection{\rqone}
To answer RQ1, we separately analyse syntactic similarity (RQ1.1), functional correctness (RQ1.2) and structural complexity (RQ1.3). In the following, we denote with \textbf{A} and \textbf{B} any two sets of values, selected from the four sets of values obtained for a given metric in the four workflows.

\subsubsection{\rqoneone} 
\begin{figure}[t!b]
    \centering
    \includegraphics[width=\linewidth]{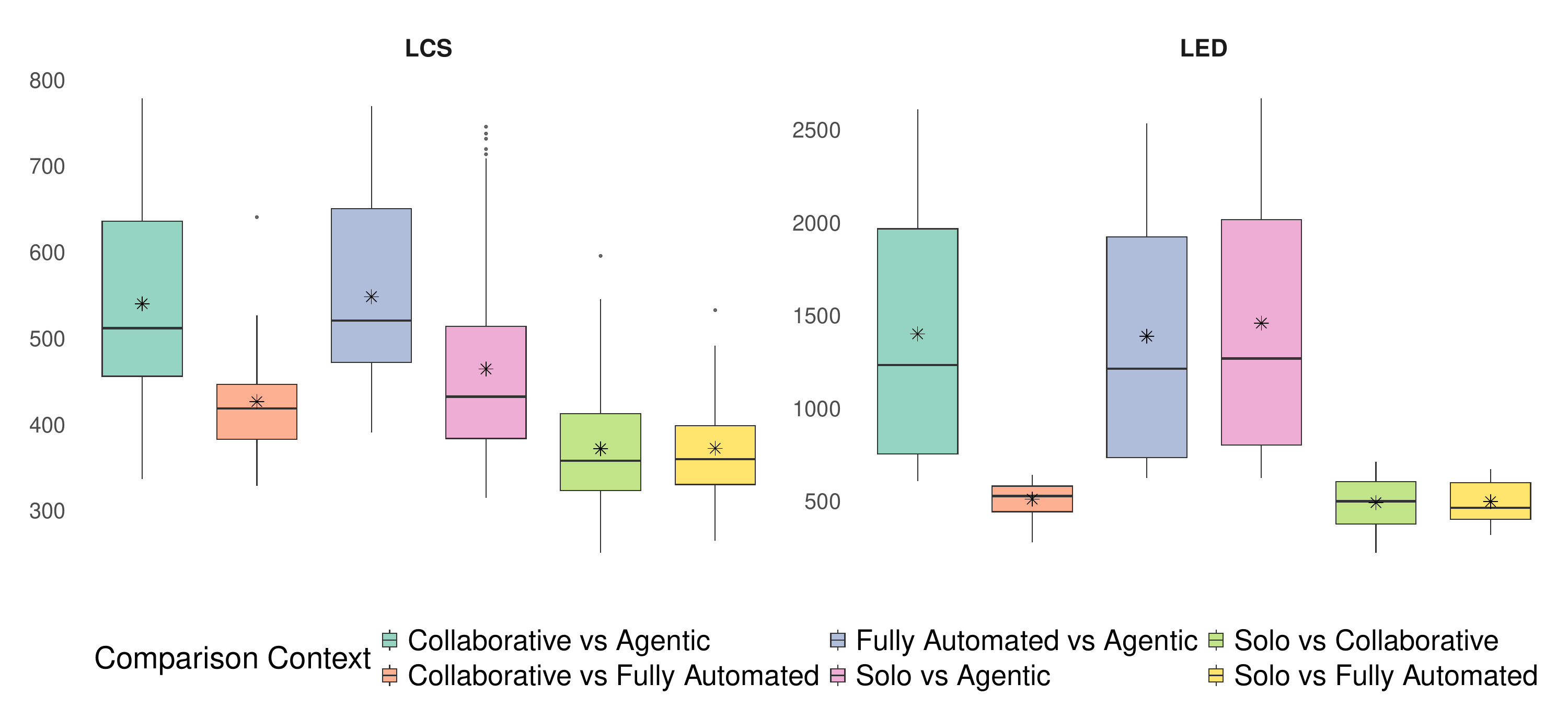}
    \caption{Syntactic similarity between any two sets A and B for \LCS and \LED. \LCS, the higher the better; \LED, the smaller the better. (*) denotes the mean value.}
    \label{fig:syntacticSimilarity}
\end{figure}

In \figref{fig:syntacticSimilarity}, we visualize the distributions of pairwise syntactic similarity of the production code measured by \LCS and \LED.
For \LCS, higher values indicate greater similarity, whereas for \LED, lower values indicate greater similarity.
The figure shows that the \agentic workflow preserves longer common code fragments with the other workflows (Large \LCS) while being locally distant (large \LED). For example, the code generated with the \agentic workflow preserves the main implementation but adds helper methods, logging, and validation.
Locally, \collaborative, \fullyautomated, and \solo workflows are very similar (low \LED) and between \solo and either of these two workflows, but they do not share large common sequences. 
Overall, the \agentic workflow preserves longer common code fragments with the other workflows while exhibiting more localized syntactic differences. In contrast, the code produced by controlling the prompts (\collaborative and \fullyautomated workflows) is locally closer, as well as closer to the code produced in \solo. Combined with the CLOC distributions in \figref{fig:boxplotsLOC}, we then summarize as in the following:
\begin{mdframed}[backgroundcolor=gray!10, nobreak=true, roundcorner=3pt]
The results indicate that the code generated by the \agentic workflow preserves long common code fragments while introducing more local modifications. In contrast, the \solo, \collaborative, and \fullyautomated workflows produce smaller implementations that differ mainly through localized edits.
\end{mdframed}
\subsubsection{\rqonetwo}
The pairwise U test in \tabref{tab:MannWhitneyRQ1TPR} reports significant differences between the \agentic workflow and the other workflows for the metric \TPR, also after HB correction,  with high effect size\footnote{Negative effect size flips the direction of the test, \ie from $A>B$ to $B>A$.}.
\begin{table}[t!b]
    \renewcommand{\arraystretch}{1.1} 
\centering
\footnotesize
\caption{One-tailed Mann–Whitney U tests for functional correctness with null hypothesis $H_0: A>B$.  (*) denote single-test statistical significance at $\alpha=0.05$. (**) denotes statistical significance after Holm-Bonferroni correction.}
\label{tab:MannWhitneyRQ1TPR}
\begin{tabular}{lccl}
\hline
\multicolumn{4}{c}{\cellcolor{gray!15}\TPR}\\ \hline
$A>B$ & U & Effect size & p (one-tailed) \\\hline
\collaborative $>$ \solo &       36.5 &        -0.16 &       \num{3.14e-01} \\
\fullyautomated $>$ \solo &       18.0 &        0.33 &        \num{2.25e-01} \\
\collaborative $>$ \agentic &        69.5 &        -0.65 &       \num{8.55e-03}** \\
\fullyautomated $>$ \agentic &         36.0 &         -1.00 &        \num{3.79e-03}** \\
\agentic $>$ \solo    &      106.5 &        0.97 &        \num{7.81e-05}** \\
\collaborative $>$ \fullyautomated &       12.0 &        -0.14 &       \num{4.08e-01} \\
\hline
\end{tabular}
\end{table}
By manually inspecting the data in~\ref{tab:experiment_result} in the appendix, we observe that the \solo and \collaborative workflows exhibit greater variability in \TPR. For example, participant P13 in the \collaborative workflow achieved a \TPR of 100.0\%, whereas participant P12 achieved only 16.7\% because the generated test code did not compile. Similarly, in the \solo workflow, participant P7 obtained the lowest \TPR (8.3\%) after creating an assertion that did not match the production logic. On average, \TPR is higher with agentic coding (94.58\%).

\begin{mdframed}[backgroundcolor=gray!10, nobreak=true, roundcorner=3pt]
The solutions produced by the \agentic workflow show the highest functional correctness, although human developers can reach the same level of correctness in \collaborative settings. Workflows involving human developers show greater variation in  functional correctness.
\end{mdframed}
\subsubsection{\rqonethree}
The pairwise U test in \tabref{tab:MannWhitneyRQ1MCC} reports significant differences between the \agentic workflow and the other workflows for the metric \MCC, also after HB correction,  with high effect size.  
It also shows significant differences between the \agentic workflow and the workflows involving humans for the metric \BCB, also after HB correction, with a high effect size. 
This indicates that the solutions produced by the \agentic workflow contain a greater number of branches that have a lower coverage of the baseline test suite than the one developed by humans.
\begin{table}[t!b]
    \renewcommand{\arraystretch}{1.1} 
\centering
\footnotesize
\caption{One-tailed Mann-Whitney U tests for code complexity with null hypothesis $H_0: A>B$. The asterisk (*) denote statistical significance at $\alpha=0.05$. (**) denotes statistical significance after Holm-Bonferroni correction.}
\label{tab:MannWhitneyRQ1MCC}
\begin{tabular}{lccl}
\hline
\multicolumn{4}{c}{\cellcolor{gray!15}\MCC}\\ \hline
$A > B$ & U & Effect size & p (one-tailed)\\\hline
\collaborative $>$ \solo &       32.5 &        0.03 &       \num{4.79e-01}  \\
\fullyautomated $>$ \solo &       11.5 &         -0.15 &       \num{6.85e-01}  \\
\collaborative $>$ \agentic &        74.5 &        -0.77 &        \num{3.13e-03}**  \\
\fullyautomated $>$ \agentic &       36.0 &         -1.00 &       \num{5.48e-03}**  \\
\agentic $>$ \solo    &      102.5 &         0.90 &        \num{2.97e-04}**  \\
\collaborative $>$ \fullyautomated &       12.0 &        0.14 &       \num{4.08e-01}  \\
\hline
\multicolumn{4}{c}{\cellcolor{gray!15}\SCB}\\ \hline
\collaborative $>$ \solo &       34.0 &        0.08 &       \num{4.15e-01} \\
\fullyautomated $>$ \solo &       12.0 &         0.11 &       \num{6.45e-01} \\
\collaborative $>$ \agentic  &       61.5 &        -0.46 &       \num{5.37e-02} \\
\fullyautomated $>$ \agentic &       12.0 &         -0.33 &       \num{8.26e-01} \\
\agentic $>$ \solo    &       82.5 &        0.53 &        \num{2.29e-02}* \\
\collaborative $>$ \fullyautomated &       11.0 &        0.05 &        \num{5.00e-01} \\ \hline

\multicolumn{4}{c}{\cellcolor{gray!15}\BCB} \\ \hline
\collaborative $>$ \solo &       25.0 &         -0.21 &       \num{7.93e-01} \\
\fullyautomated $>$ \solo &       13.0 &         -0.04 &       \num{5.87e-01} \\
\collaborative $>$ \agentic &       71.0 &        0.69 &        \num{7.55e-03}** \\
\fullyautomated $>$ \agentic &       24.0 &        0.33 &       \num{2.11e-01}\\
\agentic $>$ \solo    &       91.5 &        -0.69 &        \num{3.78e-03}** \\
\collaborative $>$ \fullyautomated &        9.0 &         0.14 &       \num{6.87e-01} \\
\hline
\end{tabular}
\end{table}
In \tabref{tab:experiment_result}, we also see that the statements and branches of the \agentic workflow solutions are covered up to at most 75\% by the baseline tests, while many participants in the \solo and \collaborative settings achieve very high branch coverage, sometimes even reaching 100\%.
For example,  participant P13 achieves a \TPR of 100.0\% with a \MCC of 8, demonstrating that high functional correctness can be achieved with relatively low structural complexity in the \collaborative workflow.

\vspace{2pt}
\begin{mdframed}[backgroundcolor=gray!10, nobreak=true, roundcorner=3pt]
The solutions produced under the \agentic workflow show more complex control-flow structures than the human-developed solutions. This increased complexity may contribute to a lower branch coverage of the baseline tests than the humans developed solutions. 
\end{mdframed}
\vspace{2pt}
\subsection{\rqtwo}
To answer RQ2, we compare the structural complexity of the test code (RQ2.1) and its code coverage with the one of the baseline suite (RQ2.2). 

\subsubsection{\rqtwoone}
To answer RQ2.1, we analyse the  structural complexity of the tests created in the workflows.
The U tests in \tabref{tab:MannWhitneyRQ2Structure} are only significant under $\alpha=0.05$ for \TAM in two pair-wise combinations involving the \agentic model with high effect size, but not significant after HB correction. 
Thus, to understand better whether the organisation of tests varies across the four groups, we qualitatively analyse the test code in the different groups, also in comparison with the organisation we derive for the baseline tests. The baseline tests are indeed organised in two test methods (one for each method to test) with five and seven assertions respectively. 
\begin{table}[t!b]
    \renewcommand{\arraystretch}{1.1} 
\centering
\footnotesize
\caption{One-tailed Mann-Whitney U tests for test structural complexity with null hypothesis $H_0: A>B$. The asterisk (*) denote statistical significance at $\alpha=0.05$. (**) denotes statistical significance after Holm-Bonferroni correction.}
\label{tab:MannWhitneyRQ2Structure}
\begin{tabular}{lccl}
\hline
$A > B$ & U & Effect size & p (one-tailed) \\\hline
\multicolumn{4}{c}{\cellcolor{gray!15}\TAM} \\ \hline
\collaborative $>$ \solo &       41.0 &         0.30 &        \num{1.54e-01} \\
\fullyautomated $>$ \solo &        7.5 &         -0.44 &       \num{9.24e-01} \\
\collaborative $>$ \agentic  &       36.5 &         -0.13 &        \num{6.97e-01} \\
\fullyautomated $>$ \agentic &       31.5 &        -0.75 &       \num{2.59e-02}* \\
\agentic $>$ \solo    &       79.0 &        0.46 &       \num{3.64e-02}* \\
\collaborative $>$ \fullyautomated &       16.5 &        0.57 &       \num{7.89e-02} \\ \hline
\multicolumn{4}{c}{\cellcolor{gray!15}\TM} \\ \hline
\collaborative $>$ \solo &       38.0 &        -0.21 &       \num{2.48e-01} \\
\fullyautomated $>$ \solo &       10.5 &         -0.22 &       \num{7.80e-01} \\
\collaborative $>$ \agentic &       52.0 &        0.24 &        \num{2.01e-01} \\
\fullyautomated $>$ \agentic &       16.5 &         -0.08 &       \num{6.24e-01} \\
\agentic $>$ \solo    &       46.5 &         -0.14 &       \num{7.27e-01} \\
\collaborative $>$ \fullyautomated &       16.5 &        0.57 &       \num{7.81e-02} \\
\hline
\end{tabular}
\end{table}
The values in \tabref{tab:experiment_result} show different testing styles, none of them perfectly matching that of the baseline tests. The \fullyautomated model is the poorest, as it consistently generates one assertion per test method and three test methods in total. The tests are rather incomplete. The tests generated with the \agentic workflow are poor in the case where the prompt is more structured (Prompt 2 in \tabref{tab:prompts}). In this case, both agents (ChatGPT and Claude) have a rather constant output with one or two assertions per method and three to six methods. In the case of a less structured prompt (Prompt 1 in \tabref{tab:prompts}), the output over multiple runs of the agents is less deterministic, also providing a method with 11 assertions. A test structure closer to the baseline tests is visible in the output of the \collaborative workflow. 
It is also worth noting that the baseline test suite does not include a test for the setter method (\texttt{setLineWidth}). Nevertheless, most human developers chose to test the setter, except for P10, P12, and P13, whose test suites result therefore structurally closer to the baseline suite. The \fullyautomated workflow also  omits the test for the setter in all its runs. In contrast, the \agentic workflow alternates between including and omitting setter tests across different runs, contributing to the observed variability in both \TAM and \TM.

Some participants (\eg P1, P10, and P11) developed comprehensive test suites covering valid, invalid, and edge cases. Others, including P2, P5, P6, and P8, focused primarily on valid and invalid inputs. We also observed attempts to improve the interaction with the CLLM through iterative test design. For example, P15 correctly identified two edge cases but, for a third one, only documented the failing scenario in the test code comments, as shown in \listref{testCaseWithComment}. Despite repeated interactions with the CLLM, the corresponding issue in the production code remained unresolved.
\vspace{2pt}

\begin{mdframed}[backgroundcolor=gray!10, nobreak=true, roundcorner=3pt]
The organization of the test suites developed in the four workflows varies. The test suites generated by the \fullyautomated and \agentic workflows are relatively less structured, whereas participants in the \collaborative workflow produced test suites whose organization is closer to the one of the baseline suite.
\end{mdframed}
\begin{table}[t!b]
    \renewcommand{\arraystretch}{1.1} 
\centering
\footnotesize
\caption{One-tailed Mann-Whitney U tests for code coverage with null hypothesis $H_0: A>B$. (*) denote statistical significance at $\alpha=0.05$. (**) denotes statistical significance after Holm-Bonferroni correction.}
\label{tab:MannWhitneyRQ2Coverage}
\begin{tabular}{lccl}
\hline
$A > B$ & U & Effect size & p (one-tailed) \\\hline
\multicolumn{4}{c}{\cellcolor{gray!15}\SC}\\ \hline
\collaborative $>$ \solo &       25.5 &         -0.19 &   \num{7.99e-01} \\ 
\fullyautomated $>$ \solo &        9.0 &         -0.33 &   \num{8.65e-01} \\ 
\collaborative $>$ \agentic &       61.0 &        0.45 &   \num{5.65e-02} \\ 
\fullyautomated $>$ \agentic &       29.0 &        0.61 &   \num{6.38e-02} \\ 
\agentic $>$ \solo    &       88.5 &        -0.64 &   \num{6.42e-03}** \\ 
\collaborative $>$ \fullyautomated &       10.5 &          0.0 &   \num{5.49e-01} \\ 
\hline
\multicolumn{4}{c}{\cellcolor{gray!15}\BC}\\ \hline
\collaborative $>$ \solo &       26.0 &         -0.17 &   \num{7.80e-01} \\ 
\fullyautomated $>$ \solo &        8.0 &         -0.41 &   \num{9.07e-01} \\ 
\collaborative $>$ \agentic &       64.5 &        0.54 &   \num{2.71e-02}* \\ 
\fullyautomated $>$ \agentic &       25.5 &        0.42 &   \num{1.48e-01} \\ 
\agentic $>$ \solo    &       87.5 &        -0.62 &   \num{7.31e-03}** \\ 
\collaborative $>$ \fullyautomated &       12.0 &        -0.14 &   \num{4.04e-01} \\ 
\hline
\end{tabular}
\end{table}
\subsubsection{\rqtwotwo}
To answer RQ2.2, we analyse the code coverage of the tests created in the workflows. After applying the HB correction, only two pairwise comparisons remain statistically significant (\tabref{tab:MannWhitneyRQ2Coverage}), corresponding to the \SC and \BC metrics. In both cases, the significant difference is between the \agentic and \solo workflows. The test suites generated in the \agentic workflow achieve lower statement and branch coverage on their corresponding generated production code than the test suites developed by participants in the \solo workflow on their own production code.  
\begin{figure}
\begin{lstlisting}[
    caption={Comment that documents which assertions fails, giving more context to the CLLM.},
    label={testCaseWithComment}
]
    # It is not working in situations where I exceed the total number of characters allowed.
    def test_words_in_the_edges(self): 
        tf = TextFormatter();
        tf.setLineWidth(20);
        self.assertEqual("foo              toy", tf.getWordsInTheEdgesOfLine("foo", "toy"))
        #self.assertEqual("pneumonoultramicrosc", tf.getWordsInTheEdgesOfLine("pneumono", "ultramicroscopic"))
        self.assertEqual("runtime             ", tf.getWordsInTheEdgesOfLine("runtime", ""))
        self.assertEqual("             runtime", tf.getWordsInTheEdgesOfLine("", "runtime"))
        self.assertEqual("                    ", tf.getWordsInTheEdgesOfLine("", ""));
\end{lstlisting}
\end{figure}
\vspace{2pt}

\begin{mdframed}[backgroundcolor=gray!10, nobreak=true, roundcorner=3pt]
Human involvement remains beneficial for achieving higher code coverage. In particular, the test suites developed in the \solo workflow achieve higher statement and branch coverage than those generated in the \agentic workflow. Moreover, enhancing the prompt used to trigger the activities of the \agentic workflow does not necessarily improve code coverage.
\end{mdframed}
\subsection{\rqthree}
\label{sec:rq3}
To answer RQ3, we analyse two process metrics:  \mIterations and \Time. For the \agentic workflow iterations were not collected. The remaining pairwise U tests are illustrated in \tabref{tab:MannWhitneyRQ3}. The table shows that \Time is significantly lower in the \agentic workflow than in any other workflow, with a large effect size in all comparisons, even after the HB correction. Furthermore, \Time in the \fullyautomated workflow is significantly lower than in the \collaborative workflow.
\begin{table}[t!b]
\renewcommand{\arraystretch}{1.1} 
\centering
\footnotesize
\caption{One-tailed Mann-Whitney U tests for process efficiency with null hypothesis $H_0: A>B$. The asterisk (*) denote statistical significance at $\alpha=0.05$. (**) denotes statistical significance after Holm-Bonferroni correction.}
\label{tab:MannWhitneyRQ3}
\begin{tabular}{lccl}
\hline
 $A > B$ & U & Effect size & p (one-tailed) \\\hline
\multicolumn{4}{c}{\cellcolor{gray!15}\mIterations} \\ \hline
\collaborative $>$ \solo &       49.0 &        -0.56 &        \num{3.57e-02}* \\
\fullyautomated $>$ \solo &        6.0 &         -0.56 &       \num{9.31e-01} \\
\collaborative $>$ \agentic &          - &            - &          -  \\
\fullyautomated $>$ \agentic &          - &            - &          -  \\
\agentic $>$ \solo    &          - &            - &          -   \\
\collaborative $>$ \fullyautomated &        8.5 &         0.19 &       \num{7.16e-01}\\ \hline
\multicolumn{4}{c}{\cellcolor{gray!15} \Time} \\ \hline
\collaborative $>$ \solo &       30.5 &         -0.03 &      \num{5.65e-01}\\
\fullyautomated $>$ \solo &       24.0 &        -0.78 &       \num{3.15e-02}*\\
\collaborative $>$ \agentic &       84.0 &        1.00 &      \num{9.37e-05}**\\
\fullyautomated $>$ \agentic &       36.0 &        1.00 &        \num{2.08e-03}**\\
\agentic $>$ \solo    &       94.5 &        -0.75 &       \num{1.28e-03}**\\
\collaborative $>$ \fullyautomated &       21.0 &         1.00 &        \num{8.97e-03}**\\
\hline
\end{tabular}
\end{table}
\tabref{tab:details_collaborative} provides additional insights into how the collaborative iterations unfold. Besides the total number of iterations, the table reports the number of successful iterations and the number of iterations requiring assertion rework, \ie when the toolset reached the maximum limit of five repetitions and returned control to the user to clarify the intended behaviour to the CLLM. On average, the \collaborative workflow required 12.1 iterations, of which 3.9 were successful and 5.6 involved assertion rework. These results indicate that a lower number of iterations does not necessarily correspond to a smoother development process, as participants frequently had to reformulate or refine their tests before the CLLM could proceed correctly. The data obtained from participants P13 and P16 illustrate this variability well. P13 completed 26 iterations, including 12 iterations for assertion rework, whereas P16 completed only four iterations, all of them required rework.
\begin{table}[t!b]
    \renewcommand{\arraystretch}{1.1} 
    \centering
    \caption{Logs of the iterations on the \collaborative model.}
    \label{tab:details_collaborative}
    \begin{tabular}{cccc}
    \hline
        ID & \mIterations & successful & with reworked assertions \\\hline
        P10 & 10 & 5 & 6 \\
        P11       &  11 & 1 & 4  \\
        P12       & 12 & 2 & 3  \\
        P13       & 26 & 9 & 12 \\
        P14       &  8 & 2 & 4  \\
        P15       & 14 & 4 & 6  \\
        P16       &  4 & 4 & 4  \\\hline 
        Mean&12.1&3.9&5.6 \\\hline 
    \end{tabular}
\end{table}
\vspace{2pt}
\begin{mdframed}[backgroundcolor=gray!10, nobreak=true, roundcorner=3pt]
The \agentic workflow achieves the shortest development time, significantly outperforming all other workflows, while the \fullyautomated workflow is also significantly faster than the \collaborative workflow. However, the analysis of collaborative iterations in the \collaborative workflow shows that development effort is not solely reflected by the number of iterations, as participants frequently had to reformulate assertions before the CLLM could proceed, leading to substantial variability across users.
\end{mdframed}

\subsection{Feedback}
\label{sec:feedback}

The post-experiment questionnaire collected: (1) how the participants perceived the task and (2) a free text field in which the participant could freely express their feedback.
\begin{table}[t!b]
    \renewcommand{\arraystretch}{1.1} 
    \centering
    \caption{Grouped post-experiment responses for exercise perception by interaction model.}
    \label{tab:feedbackExerciseScenario}
    \begin{tabular}{llll}
    \hline
         & Easy & Fine & Hard \\\hline
        \solo & P5, P6, P7, P9 & P1, P3, P4 & P2, P8 \\
        \collaborative & P10, P11, P12, P13 & P14, P15, P16 & -- \\\hline
    \end{tabular}
\end{table}
\tabref{tab:feedbackExerciseScenario} illustrates, most participants in both scenarios rated the exercise as easy or fine with the knowledge they had, while only P2 and P8 in the \solo setting reported it as hard.
At the same time, this perception does not fully align with the quantitative results: several participants who underperformed still rated the exercise as easy or fine, such as P4, P7, P9, P12, P14, and P16. 

Participants P5, P13, P14, and P16 reported a positive experience with the experiment. P5 described it as a ``Fun coding challenge'', P13 stated that ``It was fun and seems like a great idea'', and P14 commented ``I like the experiment, the tool seems easy to use''. P16 also expressed appreciation for the interaction with the tool, noting that ``The tool works, I enjoyed interacting with it'', while still observing that ``The colab platform wasn't that smooth''. In general, the positive feedback mainly concerns the engaging nature of the task and the interaction with the CLLM in the \collaborative mode. However, participant P9 noted that ``1. I was too focused on writing good comprehensive code and did not look at the time. 
2. I spent too much time on the setLineWidth function
3. The auto-complete in the Browser was much slower than expected.
''.
They were unable to complete the task within the given time frame, as they focused on producing high-quality code and consequently lost track of time. Additionally, P9 and P16 highlighted difficulties in interacting with the environment, namely, Google Colab, indicating that future studies should consider using more seamlessly integrated platforms.
Furthermore, P10 stated the following ``To be honest, the "presence" of the AI made me a little unsure in the beginning, because I was concerned about its behaviour and if I should adapt to fit its need. Once I realized the AI would adapt to my needs (in particular my dev-flow), I think the experience went way more smoothly.'' highlighting that a mental model shift needs to take place to effectively collaborate with the CLLM.

In contrast to this positive feedback, participants P8, P6, and P4 report a negative experience. P8 reported ``I didn't understand the questions but after Moritz's help it was fine'', P6 also stated ``It took me WAY LONGER to understand what the english of the exercise meant than actually do it'', and P4 expressed ``maybe providing a couple of examples strings for the tests may have speed up a bit the work and clarified some requirements''. When we analyse the correctness of their results against the baseline tests and inspect their code, we see that only the code of P4 underperformed with respect to the other participants and, in fact, misunderstood the task. The other two participants scored low on correctness, P7 and P9. P9 did not consider edge cases, while P7 misunderstood part of the task. 

The results of the survey can be obtained in the folder \texttt{responses} within the replication package \citep{MockEtAl2025journal}.
\vspace{2pt}
\begin{mdframed}[backgroundcolor=gray!10, nobreak=true, roundcorner=3pt]
Most participants enjoyed the experiment, particularly the collaborative interaction with the CLLM, although some initially struggled with the open-ended task and others faced time constraints or platform difficulties, highlighting both positive participation and areas for improvement in task design and environment integration.
\end{mdframed}
\subsection{Summary of the results}\label{sec:summary}
\rqone~The interaction model influences the quality of the generated production code. The \agentic workflow consistently produces functionally correct solutions that share long common code fragments but differ by local changes from the code produced in the other workflows, also  at the cost of greater structural complexity and lower branch coverage by the baseline test suite. In contrast, the \solo, \collaborative, and \fullyautomated workflows generate smaller and locally more similar implementations. Human involvement in \collaborative models introduces greater variability in functional correctness sometimes achieving the same correctness of code generated in the \agentic workflow.

\noindent
\rqtwo~The interaction model also affects the quality of the generated test suites. Human involvement leads to better-organized test suites that more closely resemble the baseline tests and, in the \solo workflow, achieves higher statement and branch coverage than the \agentic workflow. In contrast, the \fullyautomated and \agentic workflows produce less structured test suites, indicating that increasing prompt structure alone does not necessarily improve test quality or code coverage.

\noindent
\rqthree~The \agentic workflow is the most time-efficient, followed by the \fullyautomated workflow. In contrast, the \collaborative workflow requires frequent assertion reformulation during the interaction with the CLLM, indicating additional coordination effort despite the benefits of human involvement.
\section{Discussion}
\label{sec:discussion}
Our evaluation combines two complementary settings. The human-centred interaction models (\solo and \collaborative) were evaluated through a controlled pre-experimental study involving professional developers, whereas the \fullyautomated and \agentic workflows were evaluated through repeated executions of autonomous workflows on the same development tasks. Throughout the discussion, results are interpreted within the context of these respective evaluation settings.

The \agentic workflow consistently produces the most functionally correct production code, as measured by the baseline test suite, and is also the fastest approach. However, these advantages come with greater structural complexity and lower branch coverage when evaluated using the same test suite. One possible explanation is that some of the uncovered branches correspond to implementation decisions that are not explicitly defined by the functional specifications. This may indicate that the agents in the \agentic workflow make independent implementation decisions, introducing additional untested decision points beyond what is strictly required by the specified functionality.
This suggests that autonomous \agentic platforms can effectively solve development tasks while generating solutions that are architecturally more complex and potentially more difficult to maintain.

Human involvement remains valuable, particularly for testing. Developers may produce better-organized test suites with greater code coverage.
These findings results are compared with the repeated executions of the autonomous workflows, the exploratory evidence suggests that human judgment is especially beneficial for designing effective tests, while simply increasing prompt sophistication or delegating the process to autonomous agents does not automatically improve testing quality.
The collaborative workflow represents a compromise between automation and human expertise. It can achieve a production code correctness comparable to the \agentic workflow while producing higher-quality test suites. However, this benefit comes at the expense of efficiency, as developers frequently needed to reformulate assertions and coordinate with the CLLM, increasing the overall development effort.

The \fullyautomated workflow offers shorter development times than human-assisted approaches but does not achieve the same level of production code correctness as the agentic workflow nor the same testing quality as workflows involving humans. This suggests that prompt-driven automation alone is insufficient to match the reasoning capabilities of autonomous agentic systems or the testing skills provided by developers.

The participants' feedback complements these quantitative findings. Most developers reported a positive experience, particularly appreciating the collaborative interaction with the CLLM, indicating that such workflows are acceptable and engaging despite requiring additional effort. At the same time, reports of difficulties with the open-ended task, time pressure, and platform usability point to practical limitations that may have influenced efficiency and highlight opportunities to improve experimental environments and development tools.

\section{Threats to validity}
\label{sec:threatValidity}
The major threats we envisage in our work are related to construct, internal, and external validity.
\par\noindent\textit{Construct validity.}
Different evaluation settings. This study combines two complementary empirical settings. The \solo and \collaborative workflows were evaluated through a controlled pre-experimental study involving professional developers, while the \fullyautomated and \agentic workflows were evaluated through repeated executions of the tasks. This distinction reflects the intrinsic characteristics of the evaluated interaction models: autonomous workflows do not involve human decision making during execution and therefore cannot be assessed through the same experimental protocol adopted for human-centred workflows. To facilitate comparison, all workflows were evaluated using the same development tasks, quality metrics, and evaluation framework. Nevertheless, comparisons between human-centred and autonomous workflows should be interpreted in light of their different evaluation settings.

\par\noindent\textit{Internal validity.}
Threats in this category are related to confounding factors that affect the study’s findings. 
In our work, the results may have been affected by the experience in TDD and Python of the developers. TDD developers must have a
strong command of the practice and experience in development,~\citet{Causevic2011}. To mitigate this issue, we selected professionals with specific experience in TDD and designed two simple tasks for which basic programming knowledge was needed. 

The sample size of the experiment with humans may be considered limited.  Sampling in software engineering research studies is, in fact, an inherently difficult problem~\citep{Baltes2022} as well as recruiting professionals to participate in TDD experiments~\citep{Ghafari2020, RomanoEtAl2025}. Thus, we made sure that we achieved the typical sample size  of
studies with professionals in software engineering,~\citet{KitchenhamMadeyski2024}.
Nonetheless, the limited number of professional participants reduces the statistical power of our analyses. As increasing the sample size was not feasible, we adopted non-parametric statistical tests (Kruskal–Wallis and Mann–Whitney), which are appropriate for small samples and do not rely on normality assumptions. We also report effect sizes and apply multiple-testing correction to provide readers with a broader range of evidence on which to interpret the results.
 
The instrumentation, baseline test suite, model configuration, TDD process, and experimental environment may all have influenced the quality of the collected data and, consequently, the study results. As this work is exploratory, we acknowledge this as a relevant threat, although addressing it exhaustively was beyond the scope of the study. To facilitate future replication, we described our methodology in a way that is as independent as possible from these implementation-specific factors. Investigating the impact of variations in these factors is left for future work.

\par\noindent\textit{External validity.}
External validity concerns the ability to generalise the results. 
The development tasks we gave in our experiments are simple, but they are suitable in this exploratory study as they gave us the possibility to vertically inspect the output to interpret the quantitative tests. In addition, the tasks were  designed carefully to allow participants with different skills and experience to produce the code in a short time (40 min). 

We used a single CLLM (ChatGPT) accessed through API calls. This choice ensured a consistent experimental setting and enabled a more reliable comparison of the outputs across participants. To improve the generalisability of our findings, we also evaluated the agentic workflow using two different CLLMs, allowing us to assess whether its performance was consistent across models. 

\section{Related Work}
\label{sec:relatedWork}
There is an ongoing debate about the role of GenAI in software development. In this overview, we summarise recent research along three main directions: GenAI as a tool for code generation, as a peer collaborator, and as an enabler of the TDD process.

\noindent\textbf{GenAI as a tool.}
GenAI has the potential to significantly enhance software development productivity~\citep{Fan2023}. For example, pair programming with AI assistants can increase development speed by up to 55.8\%~\citep{PengEtAl2023SpeedGenAI}. Pre-trained LLMs are frequently fine-tuned for specific generation tasks, such as producing secure code~\citep{Li2024}, generating unit tests~\citep{Tufano2022}, or summarising code~\citep{Mastropaolo2024}. Putting attention to parameter optimisation increases the performance of fine-tuning for code generation compared to other approaches~\citep{Weyssow2025} and is particularly effective when the training data is limited~\citep{GiagnorioEtAl2025SilverBullet}.
At the same time, direct prompting pre-trained LLMs for code generation is gaining momentum~\citep{Brown2020,WangEtAl2024SurveyTestingWithLLM}. This approach is less resource-intensive than fine-tuning but may suffer from limited domain knowledge. As it does not require much background knowledge, inexperienced users may produce low-quality code without knowing it. Determining which approach (fine-tuning or prompting) is superior remains an open question, and keeping humans in the loop is generally considered advisable~\citep{ShinEtAl2025,MockEtal2024TDD}.
Regardless of the approach, the quality of the generated code is a key concern~\citep{MailachEtAl2025,ShinEtAl2025}. One strategy is to prompt LLMs iteratively and repeatedly to refactor and improve the code~\citep{FakhouryEtAl2024,White2024}. This solution requires considerable effort and may not be feasible for large, complex development projects. Our \collaborative and \fullyautomated approaches aim to address this problem by automating collaboration and prompting. 

\noindent\textbf{GenAI as a peer collaborator.} 
Vibe coding and agentic coding provide different perspectives of human-GenAI collaboration in software development, \citep{SapkotaEtAl2025VibeAgentic}. 
Much of the current research on code generation focuses on vibe coding, where GenAI acts as a full-fledged assistant with whom developers can engage in a real conversation. indeed, CLLMs have been introduced into software development to support this goal \citep{PengEtAl2023SpeedGenAI}. Without a thorough understanding of the code, conversations with CLLMs can again lead to low-quality code. 
Although vibe coding is fast and accessible, it comes with the costs of low quality (\eg technical debt), little attention to quality assurance (\eg no testing), and poor prompt design~\citep{FawzyEtAl2025}.
This situation resembles the early stages of open source development, and we believe that the strategies developed since then~\citep{ChampionHill2025} can also be effectively applied in this context.

Since 2020, agentic coding has emerged as a new frontier in software development~\citep{SapkotaEtAl2025VibeAgentic}. Agentic coding refers to the paradigm in which LLMs act as autonomous agents capable of generating  code with minimal human intervention. Building on this concept, agentic platforms coordinate multiple pre-trained LLMs (agents) that collaborate to achieve predefined software development goals. These platforms enable the selection and orchestration of different models to generate diverse code artifacts, effectively simulating an end-to-end software engineering workflow.
The LLM agents (\eg AutoGPT~\citep{autogpt},
BabyAGI~\citep{babyagi}, Devin AI~\citep{devin_ai}, OpenAI
Codex~\citep{openai_codex}, Google Jules~\citep{google_jules}) demonstrate the ability to decompose abstract
goals, generate code, invoke APIs, interact with development environments, and reason iteratively through
planning-execution-feedback loops.  
MetaGPT~\citep{HongEtAl2024metagpt} is a multi-agent platform that allows users to select the type of LLM as an agent and provides a fully automated platform for development, where the user just needs to express a development goal, and the agents act as developers working individually or in a team. All these initiatives are built on predefined development processes. For instance, MetaGPT~X relies on development workflows (Standard Operating Procedure - SOPs) that are unable to capture the iterative and incremental nature of Test-Driven Development. 
In our work, we introduced models for process development that not only automate tasks but also structure and manage interactions with CLLMs within iterative workflows. Based on our preliminary findings, we argue that the software development process itself requires rethinking. The various phases of the software life cycle may need to be redesigned to fully leverage the rapid development capabilities of CLLMs. For example, atomic design elements and functional requirements could be directly incorporated into the prompts, while quality assurance mechanisms should be redefined to automatically control the generated artefacts during the development process.

\noindent\textbf{GenAI and TDD.}
 To the best of our knowledge, three approaches are relevant for our work: LLM4TDD that develops into two subsequent studies \citep{PiyaEtAl2024TDD,PiyaEtAl2025TDD}, TGen~\citep{MathewsNagappan2024TDD4LLMCodeGeneration}, and TiCoder~\citep{FakhouryEtAl2024}. For all of them, the programming language is Python. 

Our approach focuses on integrating the collaboration with CLLMs into the TDD process and empirically assesses the effects on software quality. 
The goals of LLM4TDD and TGen somehow target the opposite problem by studying how the information coming from the tests and their execution through a TDD process can improve the ability of LLM to generate correct code. 
The goal of TiCoder is to help users select the most correct code among a list of LLM-generated snippets. 
To achieve our goal, we conducted two studies: (i) an experiment involving 16 professionals experienced in TDD, and (ii) a comparative analysis of the correctness of the code generated automatically during the experiment and the code generated using an agentic development platform. 

Instead,  LLM4TDD and TGen perform case studies mining public datasets on which their framework is applied for code generation and analysis.
The authors of TiCoder designed an interactive TDD workflow in which users prompt the LLM for a task, and the model generates a set of candidate code and test suggestions. User feedback is then incorporated to refine the suggestions for code whose tests fail. 
In TiCoder, the experiment was run with a similar number of participants, time frame for the experiment run and with an additional pre-experiment trial to ours. However, it was conducted with nine PhD students and six professionals from Microsoft and the pre-experiment trial involved three professionals. 
We follow a low-granularity version of TDD, in which each TDD iteration executes one assertion at a time, \ie assertion-first. Both LLM4TDD and TGen perform TDD iterations based on failing tests, which may include multiple assertions per test case. In TiCoder, LLM generates tests and code based on initial tasks. The iteration is performed to incorporate user's feedback. 
To evaluate the accuracy of the respective approach, our work, LLM4TDD and TGen define a set of tests as ground truth. 
In LLM4TDD, tests are developed manually using partition, boundary, and state-based testing and generated with LLM. TGen, on the other hand, relies entirely on existing unit tests developed by humans from MBPP and HumanEval, plus synthetically generated ones from EvalPlus.  
In TiCoder, tests are generated by the LLM and refined with the user's feedback. Also in this case, the iteration is triggered when some tests fail. 
In addition, TiCoder uses the MBPP and HumanEval tests to evaluate the accuracy of the generated code. 
Our baseline tests are the ground truth for all our experiments. They are crafted manually through category partition testing and are derived from specifications. 

LLM4TDD counts the number of prompts against the number of ground truth tests and correctness. TGen focuses primarily on correctness using the percentage of passed tests in different research problem settings. TiCoder uses time and cognitive load to measure the performance of developers in evaluating AI output and pass@k of the ground truth to measure the accuracy of the generated code.
LLM4TDD and TGen find that more than 80\% of the problems can be solved without reiterating with prompts and hinting with human intervention or more information on the test and their execution. TiCoder finds that incorporating users' intent boosts LLM performance to generate correct code. 
These results are based on existing datasets of problems and a limited number of tests (up to 4 for MBBP and HumanEval with TGen), tests and iteration (for LLM4TDD, up to 10 cases and iterations) or tasks and related tests (for TiCoder). Our work does not set limits on iterations or in the process of generating the baseline tests. Our qualitative analysis also suggests that intensive collaboration with a CLLM can produce functionally correct code,  performing as well as an \agentic platform, although the \agentic workflow statistically produces higher-quality code.

\section{Conclusions and Future Work}
\label{sec:conclusion}
This study presents an exploratory investigation of human–AI collaboration in software development. We implemented and evaluated different interaction models, assessing their impact on the quality of the resulting software artifacts. We conducted a controlled pre-experimental study involving professional developers to evaluate the \solo and \collaborative workflows. Then, we replicated the study with \fullyautomated and \agentic workflows to provide exploratory evidence on autonomous software generation. To compare the output resulting from the four models of interaction,  we developed an evaluation framework  to assess quality of code and process efficiency.

Our findings indicate that the choice of an interaction model should depend on the development objective. Agentic workflows are particularly suitable when rapid development and production code correctness are the primary goals. However, they may also introduce additional implementation decisions that are not explicitly required by the functional specifications, resulting in untested decision points.
Conversely, workflows involving humans remain preferable when high-quality testing, better-organized test code, and greater control over the development process are required. Collaborative approaches offer a promising balance between these objectives but require improved interaction mechanisms to reduce the coordination overhead between developers and CLLMs.

In future work, we plan to extend our investigation to additional development practices, CLLM, and agentic platforms, while also exploring how AI may reshape the software development process itself. We aim to investigate whether traditional phase-based and iterative development models remain appropriate in an era where AI systems can generate complete software solutions without explicitly progressing through conventional phases such as design.

We also plan to apply our \fullyautomated evaluation suite to established benchmarks such as MBPP~\citep{AustinEtAl2021MBPP} and HumanEval~\citep{ChenEtAL2021HumanEval}, which provide problem specifications together with reference implementations and test suites, to further assess the quality of the generated software artifacts.

\section*{Acknowledgement}
The research has been funded by the European Commission Next Generation EU as part of the project PRIN2022 2022TEPX4R - Behaviour-enabled IoT (BeT), CUP 53D23003730008 and the European Union- Next Generation EU, Mission 4 Component 1 CUP I52B23000570003.
\newpage

\section*{Appendix}
\begin{table}[ht]
    \renewcommand{\arraystretch}{1.1} 
    \centering
    \footnotesize
 \setlength{\tabcolsep}{4pt}
    \renewcommand{\arraystretch}{0.95}
    \caption{Metrics' values per participant (vibe coding) and per run (agentic coding) and means and standard deviations.  The brackets for the values  of \TAM  indicate the total number of assertions. (*) Participants perceiving time $>$40 min (actual limit: 40 min). Abbreviations: CG (ChatGPT), CS (Claude Sonnet), FA (\fullyautomated).}
    \label{tab:experiment_result}
    \begin{tabular}{C{0.07\textwidth}|C{0.07\textwidth}cC{0.07\textwidth}C{0.07\textwidth}|C{0.12\textwidth}cC{0.07\textwidth}C{0.07\textwidth}|cC{0.07\textwidth}}
    \hline
    ~ & \multicolumn{4}{c|}{Production code} & \multicolumn{4}{c|}{Test code} & \multicolumn{2}{c}{Process}\\
    \hline
    ID & \TPR ($\%$) & \MCC & \SCB (\%) & \BCB (\%) & \TAM(Ass.)  & \TM & \SC (\%) & \BC (\%) & Iter. & \Time (sec) \\\hline
    \multicolumn{11}{c}{\cellcolor{gray!15}\solo } \\\hline
    P1 & 83.3 & 9 & 65 & 88 & 1.55 (17) & 11 & 100 & 100 & 26 & 2,100  \\
    P2 & 83.3 & 5 & 64 & 100 & 1 (3) & 3 & 100 & 100 & 13 & 2,400  \\
    P3 & 75.0 & 6 & 64 & 100 & 1.5 (6) & 4 & 91 & 75 & 22 & 1,200  \\
    P4 & 33.3 & 7 & 53 & 100 & 3 (3) & 1 & 67 & 25 & 25 & 2,100  \\
    P5 & 83.3 & 4 & 64 & 100 & 1 (3) & 3 & 100 & 100 & 12 & 2,400  \\
    P6* & 75.0 & 8 & 14 & 0 & 1 (3) & 3 & 100 & 100 & 22 & 4,200 \\
    P7 & 8.3 & 7 & 45 & 100 & 1 (3) & 3 & 100 & 100 &  9 & 600  \\
    P8* & 75.0 & 5 & 54 & 100 & 1 (3) & 3 & 100 & 100 & 21 & 3,600  \\
    P9 & 41.7 & 7 & 53 & 67 & 1.27 (14) & 11 & 100 & 100 & 29 & 1,800  \\
    \hline
    Mean & 62.0 & 6.44 & 52.89 & 83.89 & 1.37 (6.11) & 4.67 & 95.33 & 88.89 & 19.89 & 2,266 \\
    SD & 27.3 & 1.59 & 16.17 & 33.35 & 0.65 (5.47) & 3.67 & 11.03 & 25.34 & 6.94 & 1,103\\
    \hline
    \multicolumn{11}{c}{\cellcolor{gray!15}\collaborative} \\\hline
    P10 & 91.7 & 9 & 64 & 83 & 1 (7) & 7 & 91 & 83 & 10 & 1,800 \\
    P11 & 83.3 & 4 & 42 & 100 & 3 (9) & 3 & 100 & 100 & 11 & 1,800 \\
    P12 & 16.7 & 4 & 67 & 100 & 1 (4) & 4 & 100 & 100 & 12 & 2,400  \\
    \textbf{P13} & \textbf{100.0} & \textbf{8} & \textbf{62} & \textbf{83} & \textbf{6 (18)} & \textbf{3} & \textbf{100} & \textbf{100} & \textbf{26} & \textbf{2,400}  \\
    P14 & 33.3 & 6 & 64 & 75 & 1.6 (8) & 5 & 86 & 50 & 8 & 2,400  \\
    P15 & 83.3 & 6 & 56 & 100 & 2.67 (8) & 3 & 100 & 100 & 14 & 2,400  \\
    P16 & 16.7 & 9 & 31 & 12 & 1 (5) & 5 & 85 & 62 & 4 & 1,200  \\\hline
    Mean & 60.8 & 6.57 & 55.14 & 79.00 & 2.32 (8.43) & 4.29 & 94.57 & 85.00 & 12.14 & 2,057 \\
    SD & 36.8 & 2.15 & 13.55 & 31.25 & 1.82 (4.58) & 1.50 & 7.02 & 21.05 & 6.89 & 472 \\
    \hline
    \multicolumn{11}{c}{\cellcolor{gray!15}\fullyautomated}  \\\hline
    FA1 & 83.3 & 4 & 36 & 100 & 1 (3) & 3 & 100 & 100 & 4 & 720 \\
    FA2 & 83.3 & 7 & 50 & 33 & 1 (3) & 3 & 94 & 67 & 18 & 960 \\
    FA3 & 83.3 & 7 & 77 & 100 & 1 (3) & 3 & 91 & 75 & 16 & 720 \\\hline
    Mean & 83.3 & 6.0 & 54.33 &  77.67 & 1 (3) & 3 & 95.00 & 80.67 & 12.67 & 800 \\
    SD & 0.0 & 1.73 & 20.84 & 38.68 & 0 (0) & 0 & 4.58 & 17.21 & 7.57 & 138 \\
    \hline
    \multicolumn{11}{c}{\cellcolor{gray!15}\agentic}  \\\hline
    CG1\_1 & 83.0 & 17 & 68 & 62 & 1.44 (13) & 9 & 100 & 100 & - & 120 \\
    CG1\_2 & 100.0 & 15 & 64 & 62 & 11 (11) & 1 & 97 & 94 & - & 120 \\
    CG1\_3 & 100.0 & 13 & 58 & 58 & 10 (10) & 1 & 94 & 83 & - & 120 \\
    CG2\_1 & 92.0 & 12 & 60 & 60 & 1 (3) & 3 & 80 & 50 & - & 60 \\
    CG2\_2 & 92.0 & 9 & 62 & 67 & 1 (3) & 3 & 88 & 50 & - & 60 \\
    CG2\_3 & 92.0 & 8 & 61 & 75 & 1 (3) & 3 & 89 & 50 & - & 60 \\
    CS1\_1 & 100.0 & 13 & 70 & 71 & 1.5 (15) & 10 & 93 & 86 & - & 60 \\
    CS1\_2 & 100.0 & 16 & 52 & 60 & 1.8 (18) & 10 & 71 & 80 & - & 60 \\
    CS1\_3 & 92.0 & 12 & 71 & 70 & 14 (14) & 1 & 86 & 60 & - & 60 \\
    CS2\_1 & 100.0 & 9 & 75 & 75 & 2 (3) & 6 & 86 & 50 & - & 60 \\
    CS2\_2 & 92.0 & 8 & 69 & 62 & 2 (8) & 4 & 85 & 50 & - & 60 \\
    CS2\_3 & 92.0 & 8 & 69 & 62 & 2 (6) & 3 & 86 & 50 & - & 60 \\\hline
    Mean & 94.58 & 11.67 & 64.92 & 65.33 & 4.06 (9.17) & 4.25 & 87.92 & 66.92 & - & 75.0 \\
    SD & 5.40 & 3.26 & 6.54 & 6.02 & 4.69 (5.13) & 3.41 & 7.75 & 19.97 & - & 27.14 \\\hline
    \end{tabular}
\end{table}
\newpage

\bibliographystyle{spbasic}   
\bibliography{ref.bib}
\end{document}